\documentclass[journal=ancac3,manuscript=article,super=true]{achemso}
\usepackage{graphicx}
\usepackage{amsmath,amsfonts,amssymb,mathtools,empheq} % Math packages
\usepackage{textcomp}
\usepackage{gensymb}
\usepackage{siunitx}
\usepackage{caption}
\usepackage{subcaption}
\usepackage{float}

\usepackage[normalem]{ulem} % For underlining text elements
\usepackage{bm} % For bolding symbols

\mciteErrorOnUnknownfalse

\usepackage{hyperref}
\usepackage{orcidlink}

\newcommand*{\AaltoAffiliation}{Department of Applied Physics, Aalto University School of Science, P.O. Box 15100, Aalto, FI-00076, Finland}

\newcommand*{\TurkuAffiliation}{Department of Mechanical and Materials Engineering, University of Turku, FI-20014 Turku, Finland}

\newcommand*{\KTUAffiliation}{Faculty of Chemical Technology
Department of Polymer Chemistry and Technology, Kaunas University of Technology, LT-51423 Kaunas, Lithuania}

\newcommand*{\TUeAffiliation}{Department of Applied Physics, Eindhoven University of Technology, 513, 5600 MB Eindhoven, Netherlands}

\newcommand*{\ExeterAffiliation}{Department of Physics and Astronomy, University of Exeter, EX4 4QL, Exeter, United Kingdom}

\author{Aron J. J. Dahlberg\,\orcidlink{0009-0000-9083-1729}}
\affiliation{\AaltoAffiliation}
\altaffiliation{A.D, M.G, and M.M contributed equally to this work.}
\author{Matas Gužauskas\,\orcidlink{0000-0001-6603-5298}}
\affiliation{\KTUAffiliation}
\altaffiliation{A.D, M.G, and M.M contributed equally to this work.}
\author{Malek Mahmoudi\,\orcidlink{0000-0002-9580-6220}}
\affiliation{\TurkuAffiliation}
\altaffiliation{A.D, M.G, and M.M contributed equally to this work.}

\author{Joel Lehikoinen\,\orcidlink{0000-0001-8727-9822}}
\affiliation{\AaltoAffiliation}
\author{Dmytro Volyniuk\,\orcidlink{0000-0003-3526-2679}}
\affiliation{\KTUAffiliation}
\author{Manish Kumar\,\orcidlink{0000-0001-5510-9634}}
\affiliation{\TurkuAffiliation}
\author{Anton Matthijs Berghuis\,\orcidlink{0000-0002-1896-7119}}
\affiliation{\TUeAffiliation}
\author{Hassan A. Qureshi\,\orcidlink{0000-0002-9065-2525}}
\affiliation{\TurkuAffiliation}
\author{Rasa Keruckiene\,\orcidlink{0000-0002-9809-5815}}
\affiliation{\KTUAffiliation}
\author{Michael A. Papachatzakis\,\orcidlink{0000-0001-6466-1543}}
\affiliation{\TurkuAffiliation}
\author{Evgeny A. Mamonov\,\orcidlink{0000-0001-6550-8962}}
\affiliation{\AaltoAffiliation}
\author{Rebecca Heilmann\,\orcidlink{0000-0003-2716-2571}}
\affiliation{\AaltoAffiliation}
\author{Henri Lyyra\,\orcidlink{0000-0002-9218-4657}}
\affiliation{\TurkuAffiliation}
\author{Olli Siltanen\,\orcidlink{0000-0002-7295-2065}}
\affiliation{\TurkuAffiliation}

\author{William L. Barnes\,\orcidlink{0000-0002-9474-5534}}
\affiliation{\ExeterAffiliation}
\author{Jaime Gómez Rivas\,\orcidlink{0000-0002-1103-9598}}
\affiliation{\TUeAffiliation}

\author{Konstantinos S. Daskalakis\,\orcidlink{0000-0002-3996-5219}}
\affiliation{\TurkuAffiliation}
\email{konstantinos.daskalakis@utu.fi}
\author{Juozas Vidas Gražulevičius\,\orcidlink{0000-0002-4408-9727}}
\affiliation{\KTUAffiliation}
\email{juozas.grazulevicius@ktu.lt}
\author{Päivi Törmä\,\orcidlink{0000-0003-0979-9894}}
\affiliation{\AaltoAffiliation}
\email{paivi.torma@aalto.fi}

\title{Nanoparticle Arrays for Efficient Organic Light-Emitting Diode Emission Management}

\begin{document}

\begin{abstract}
%Abstract roughly 200 words. Check specific journal guidelines once decided. 
OLEDs are increasingly applied in illumination and displays because they offer excellent color quality, are mechanically flexible, and are self-emissive. However, their usage is limited by low external quantum efficiency (EQE) and efficiency roll-off at high driving voltages. These limitations, together with demands for smaller pixels and device sizes in emerging technologies, motivate innovations that increase efficiency and allow replacing external optical elements with embedded solutions. Here, we demonstrate enhanced outcoupling as well as directional and polarization control of OLED emission, based on collective surface lattice resonances of plasmonic nanoparticle arrays that are embedded in the active layers of four different state-of-the-art OLED structures. Both square arrays and more complex lattices producing flat bands are demonstrated to guide the light to directions and polarizations determined by their optical modes. We show that by the design of the array geometry and the OLED structure, spectral and angular enhancement of the electroluminescence (EL), up to 30~\%, can be achieved. Our results verify that surface lattice resonances of nanoparticle arrays offer a robust and versatile embedded solution for tailoring the OLED emission, as well as exciting prospects for efficiency increase if combined with narrow-spectrum emitters. 

\vspace{\baselineskip}
\end{abstract}

% No more than 1000 words, check specific requirements later
\newpage
\section*{Introduction}

Organic light-emitting diodes (OLEDs)~\cite{Tang1987} have become a key technology due to their excellent color quality, compatibility with flexible substrates, and thin, self-emissive device architecture. While emitters with near-unity internal quantum efficiencies~\cite{Adachi2001, Wu2018} exist and emitters with narrow emission spectra~\cite{Hatakeyama2016, Chan2021} have recently been introduced, the performance of conventional OLEDs remains constrained. Significant outcoupling losses within the stack limit the external quantum efficiency (EQE) of conventional OLEDs to $\approx$20–30~\%~\cite{Gather2015, Kang2024}, polarizing secondary optics in display applications further reduce the fraction of usable light by up to 50~\% for unpolarized emission~\cite{ding2023recent}, and the commonly used emitters exhibit broad spectra. 

The magnitude of the in-plane component of the wavevector $k_\parallel$ of emitted light determines whether the emitted light can escape the OLED stack. The in-plane component $k_\parallel$ is related to the propagation angle $\theta$ via $k_\parallel = k_0 \sin \theta$, where $k_0$ is the free-space wavenumber. Figure~\ref{fig:1}a shows how different values of $k_\parallel$ correspond to the air mode and the loss channels (excluding material absorption): total internal reflection at the substrate--air (substrate modes) and internal layer (waveguide modes) interfaces, and excitation of surface plasmon polaritons (SPPs) at the metallic electrode interface.

External solutions, such as microlens arrays~\cite{Moller2002}, substrate surface roughening~\cite{Zhou2011, Riedel2010}, and high-index substrates that further employ one of the former methods~\cite{Gaertner2008, Reineke2009} have been shown to enhance the efficiency of OLEDs. However, such solutions can lead to incompatibilities with form factor requirements and added complexity in fabrication. In turn, embedded solutions such as microstructuring of the internal layers~\cite{Lupton2000, Youn2015}, low-index grids~\cite{Sun2008}, and scattering layers~\cite{Kim2018, Song2018, Jeon2018, Zhao2024} have been shown to extract light from internal loss pathways. These often do not provide control over directionality or polarization. Metasurfaces have been applied for directionality and polarization of the emitted light along with engineering of the intrinsic dipole orientation of the emitters~\cite{Zeng2024, ding2023recent}. Recently, control of both was obtained by structuring the OLED pixels themselves through nanostencil fabrication~\cite{Marcato2025}. Optically engineered narrow and spectrally flat emission has been demonstrated in metal-clad microcavity OLEDs operating in the strong light--matter coupling regime (polariton OLEDs), but such architectures present challenges for managing outcoupling losses~\cite{Gubbin2014, Mischok2023, Abdelmagid2024}.

From the aforementioned solutions, a need for versatile embedded solutions that can enhance outcoupling while enabling control over the angular, polarization, and spectral characteristics is recognized. In systems such as augmented/virtual reality (AR/VR), these characteristics must be achieved with miniaturized light sources, for which embedded solutions are a necessity. Embedded nanoparticle arrays in square-like geometries have been shown to enhance current efficiency and luminance, and change the color of OLEDs~\cite{Auer-Berger2017-1, Auer-Berger2017-2, Auer-Berger2022}. The effects of the nanoparticles on the device's EQE and EL polarization have not been quantitatively analysed to date. Furthermore, the effects depend on the OLED device design as well as the array geometry. A systematic study linking OLED waveguide modes, emitter spectrum, array geometry, and the resulting effects on emission properties of OLEDs has not yet been performed. 

In this manuscript, we study embedded plasmonic nanoparticle arrays inside various high-performing OLEDs. The arrays host collective modes called surface lattice resonances (SLRs), whose spectral position, angular dispersion, and polarization dependence can be tuned through the lattice geometry~\cite{Zou2005, Auguie2008, Vecchi2009, wang2018rich, Kravets2018}. In the OLED stack, SLRs provide diffractive outcoupling of the OLED waveguide modes by changing the in-plane wavevector (see Fig.~\ref{fig:1}a). Here, we provide design principles for matching the SLR modes to the OLED waveguide modes and emitter spectrum. We study both square and flat band lattice geometries across four state-of-the-art device designs: a single-layer (SL) CzDBA-based device~\cite{kotadiya2019efficient}, and multilayer devices based on DMAC-DPS (DD)~\cite{zhang2015nearly}, DACT-II (D2)~\cite{Kaji2015-ic}, and DQBC (DQ)~\cite{https://doi.org/10.1002/adma.202103293}. Hereafter, we refer to the devices using the abbreviations given in parentheses above. The suffix ``NAROLED" refers to a device with an embedded nanoparticle array, ``REF" to a reference pixel (without nanoparticles) on the same substrate, and ``pure reference" to a pixel on a substrate that has not undergone the nanofabrication processing. We measured the optoelectrical properties (current-voltage-luminance), angle-resolved EL spectra, and EQEs of the devices for different nanoparticle array periods. The main contribution of our work is to establish a transferable framework for embedded nanoparticle array mode engineering in OLEDs. With simple design principles, we show that the emission near the wavelengths of the array modes and near-normal directions can be enhanced. Here, a mode enhancement of 1.3 and a degree of linear polarization of 0.3 were obtained. Furthermore, an EQE enhancement factor of up to 1.46 at low drive current was obtained. These results establish embedded nanoparticle arrays as a viable framework for SLR-based mode engineering, with control of directionality, spectrum, polarization, and EQE that can be tuned through geometry-dependent coupling.

\section*{Results}

\begin{figure}[H]
    \centering
    \includegraphics[width=0.95\linewidth]{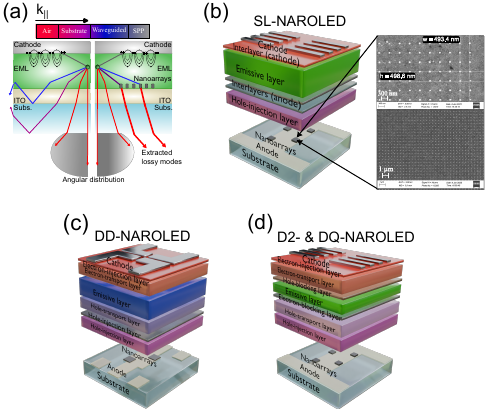}
    \caption{(a) Schematic portraying the loss paths (SPP-coupling, waveguiding and substrate modes), outcoupling and angular distribution from a typical OLED (left) and NAROLED (right) devices determined by the in-plane wavevector $k_{||}$. The nanostructures can break the interface symmetry and lead to outcoupling of otherwise lost modes as well as modify the angular distribution of the outcoupled modes. (b) SL-NAROLED with the nanoarrays at specific pixel locations. The inset shows SEM  images of a square array of aluminium nanoparticles with a period of 500~nm and a diameter of 100~nm. (c) Schematic diagram of the DD-NAROLED. (d) Schematic diagram of the D2- and DQ-NAROLEDs.}
\label{fig:1}
\end{figure}

A simplified schematic of the effects NAROLEDs have on the emission properties of the OLED is presented in Figure~\ref{fig:1}a. On the left, a normal OLED shows the outcoupled and lossy modes (SPP-coupled, waveguide, and substrate) originating from various emission angles. On the right side, a NAROLED device schematic presents how an embedded nanostructure within the stack can outcouple these modes. The SLR modes couple to the waveguided emission within the OLED stack, leading to the extraction of otherwise trapped light and enhancement of the outcoupling efficiency. A visual indication of this outcoupling mechanism is shown in Figure~\ref{fig:SI-InactivePixels}, where light is seen emitted from an inactive NAROLED pixel when another pixel in the sample is activated. Furthermore, the nanostructures provide an internal embedded scheme for modifying the emission characteristics, including angular distribution, spectral profile, and polarization.

The OLED devices used in this study all employed a bottom-emitting architecture, with the corresponding OLED stack schematics presented in Figures~\ref{fig:1}(b--d). The description of each device and the constituent layer materials is provided in the Methods section. In all devices, aluminium nanostructures were embedded within the device stack and positioned on top of the transparent ITO-anode. Their quality was examined via scatterometry, with retrieved dimensions of particle radius, particle height, and ITO thickness obtained as fitted parameters to the data. The homogeneity was found to be excellent, as shown in Figure~\ref{fig:SI-Sample_Metrology}. The insets of Figure~\ref{fig:1}b show scanning-electron microscope (SEM) images of a metallic nanoparticle array with a period of 500~nm and a diameter of 100~nm arranged in a square geometry on top of an ITO-coated substrate, representing the NAROLED device structure. All devices contained pixels with a size of 4~mm$^2$ and array sizes ranging from 0.25--4~mm$^2$.

The SL-NAROLED of Figure~\ref{fig:1}b was selected as one of the studied configurations, as it enables the nanoparticles to be positioned in proximity to the emissive layer (CzDBA). In the article introducing the corresponding SL-REF device, it was reported to exhibit high electrical efficiency, low operational voltage, ambient stability, and high EQE~\cite{kotadiya2019efficient}. The SL-NAROLED, therefore, permits an assessment of potential changes in electrical performance and whether reducing the separation between the emitters and the nanoparticles improves outcoupling.

The DD-NAROLED device shown in Figure~\ref{fig:1}c was selected to investigate the impact of nanoparticle arrays in efficient blue-emitting OLEDs. The blue thermally activated delayed fluorescence (TADF) emitter DMAC-DPS exhibits nearly concentration-independent behavior, allowing it to maintain high optical quality without aggregation~\cite{zhang2015nearly,Zhang2014}. The corresponding DD-REF devices have been previously optimized~\cite{Kumar2025}. Additionally, aluminum nanoparticles exhibit favorable plasmonic properties in the blue spectral region, including lower losses compared to longer wavelengths~\cite{west2010searching}.

To further evaluate the impact of nanoparticle arrays on highly optimized state-of-the-art devices, the D2-REF and DQ-REF devices were studied. The former serves as a benchmark for high-efficiency TADF performance, with the emitter DACT-II previously demonstrated to achieve nearly 100~\% internal quantum efficiency (IQE)~\cite{https://doi.org/10.1002/adma.202103293}. The corresponding D2-NAROLED facilitates the study of nanoparticle array interactions in a system where the primary limitation is light outcoupling rather than internal exciton generation. The DQ-REF device was selected due to the emitter (DQBC) exhibiting a strong horizontal dipole orientation of approximately 92~\% and a high photoluminescence quantum yield of 95~\%~\cite{https://doi.org/10.1002/adma.202103293}. Horizontal dipoles inherently improve light extraction compared to isotropic emitters, and the DQ-NAROLED allows for an assessment of whether nanoparticle arrays can provide additional enhancement to devices that are already optically optimized. A schematic for both of the devices is shown in Figure~\ref{fig:1}d.

\subsection*{Device stability}
With the inclusion of metallic nanoparticles within an electrically driven device, questions about its operational stability naturally arise. The electrical properties were characterized by studying the current density-voltage (JV) behaviour of the devices. The JV curves in both logarithmic and linear scales are found in the Figure~\ref{fig:SI - JV Curves}. The SL-NAROLED and SL-REF are compared to a pure reference in Figure~\ref{fig:SI - JV Curves}(a). There, the pure reference and the SL-REF behave similarly throughout the JV-sweeps, indicating that the nanostructure fabrication does not compromise the pixel quality. Additionally, the SL-NAROLED pixels operate similarly to reference pixels at low voltages and in the injection-limited region, suggesting that the energy-level alignment within the OLED stack remains unchanged~\cite{li2021interface,kanatsiopoulos2025correlation}. At higher voltages, however, the NAROLEDs' current density starts to deviate from the references. The deviation is seen better from the linear scale inset. The nanoparticle array size does not appear to significantly influence this behaviour, as increasing the array area by a factor of 16 did not further reduce the current density. Each JV measurement was repeated over five cycles, and the standard error was calculated from these results. In each pixel, the standard error is only visible in the data below the turn-on point, indicating that the pixels are stable over repeated measurements.

The electrical characteristics of the DD-REF and DD-NAROLED devices shown in Figure~\ref{fig:SI - JV Curves}(b) were recorded over three independent JV-measurement cycles to verify reproducibility and assess degradation effects. The DD-REF exhibits only slightly higher current density than the DD-NAROLED. A spin-coated hole-injection-layer (HIL) (PEDOT:PSS) was found to be crucial for the DD-NAROLED, as omitting this layer resulted in instability and shorting. The nanostructures with a height of 30~nm introduce topographical irregularities into the subsequently vapor-deposited OLED layers. The spin-coated HIL smoothens the surface and simultaneously improves charge injection by better aligning the energy levels of the anode with the active layers of the OLED.
 
The D2- and DQ-NAROLED devices were also found to have similar electrical characteristics to the reference devices. Their JV-curves will be presented in the context of Figure~\ref{fig:4}.

\subsection*{Directional Electroluminescence}
The control of angular distribution and spectral profile of the emission was investigated with angle-resolved detection via Fourier-plane imaging and goniometric measurements of the EL emission from OLED pixels. The intensity was normalized relative to the corresponding reference device, according to
\begin{equation}
    \label{e:reduced_intensity}
    I_{\mathrm{reduced}}=\frac{I_{\mathrm{sample}}-I_{\mathrm{ref}}}{\mathrm{max}     (I_{\mathrm{ref}})}.
\end{equation}

\begin{figure}[H]
\includegraphics[width=\textwidth]{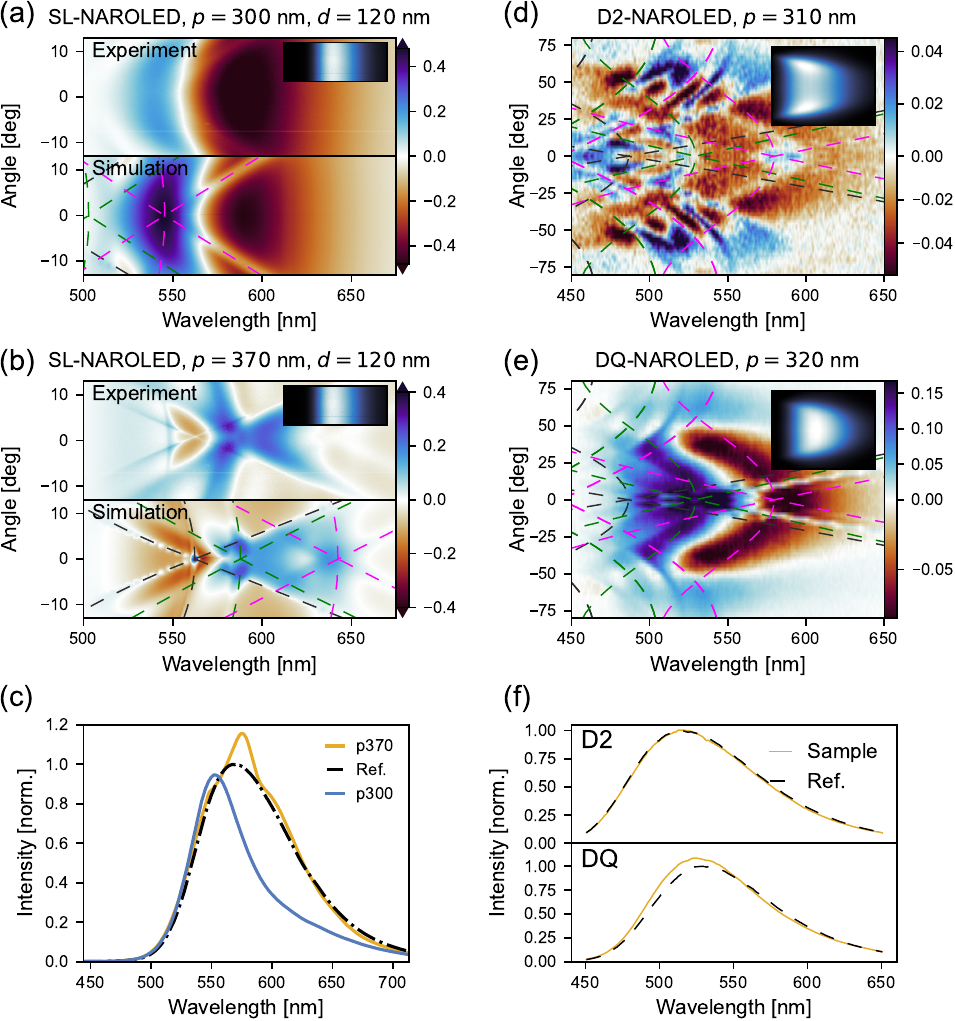}
\caption{(a, b) Angle-resolved EL taken at \SI{125}{\milli\ampere\per\centi\meter\squared} (top) and simulated emission pattern (bottom) of SL-NAROLED with a square array with periods $p$ as indicated in panel titles. (c) Integrated spectra (between $\pm 13\degree$) of the devices of panels (a) and (b). (d, e) Angle-resolved EL taken at \SI{20}{\milli\ampere\per\centi\meter\squared} from D2- and DQ-NAROLEDs with square arrays. In panels (a), (b), (d), and (e), the color scale shows the reduced intensity of Eq.~\eqref{e:reduced_intensity}, where blue means enhancement and brown de-enhancement; the insets show the EL of the reference devices. Dashed black, green, and magenta lines in the plots show the positions of DOs propagating with the refractive indices of the TM$_1$, TE$_0$, and TM$_0$ waveguide modes (see Methods for details), respectively.   (f) Angle-integrated EL spectra of the D2 (top) and DQ (bottom) devices.} 
\label{fig:2}
\end{figure}

The observed modes comprise hybridization between the OLED waveguide modes and the SLR modes, known in the literature as quasi-guided modes, guided-mode resonance, or hybrid plasmonic-photonic modes. All OLEDs studied here support three waveguide modes: TM$_0$, TE$_0$, and TM$_1$, in the order of decreasing effective index (see Fig.~\ref{fig:effective_indices} in Methods). Following this, the hybrid modes will be named as TM$_{1}^{m_1,m_2}$, TE$_{0}^{m_1,m_2}$, and TM$_{0}^{m_1,m_2}$, where the base and the subscript indicate the waveguide mode polarization and order, respectively. The superscript identifies the diffracted order (DO) $(m_1, m_2) \in \mathbb{Z}^2$ of the SLR mode; here $m_1$ ($m_2$) corresponds to diffraction out of (in) plane of incidence (the propagation axis of the in-plane wavevector): in our notation, $m_1$ ($m_2$) is associated with $k_x$ ($k_y$). The angle in all our results is given by the in-plane wavevector $k_y$; this plane of incidence is chosen in the experiments by the orientation of the spectrometer slit. Parallel and perpendicular polarizations are defined with respect to the $k_y$ axis. The refractive indices, in which the hybrid modes propagate, are adopted from those of the waveguide modes. In Figures~\ref{fig:2} and~\ref{fig:3}, the curves of the DOs in the refractive indices of the TM$_1$ (black dashed), TE$_0$ (green dashed), and TM$_0$ (magenta dashed) modes calculated using Eq.~\eqref{e:dispersion_rectangular} in Methods are shown. We associate features in the experimental and simulated data with specific quasi-guided modes by visual comparison. For further details on the topic, see Supporting Information.

Figures~\ref{fig:2}(a) and~\ref{fig:2}(b) present the angle-resolved EL spectra of SL-NAROLED pixels containing nanostructures in a square lattice geometry with periods of 370~nm and 300~nm, respectively. The spectra were measured at a constant current-density of~\SI{125}{\milli\ampere\per\centi\meter\squared} using Fourier-plane imaging.  Furthermore, to compare the observations to predictions from theory, far-field emission patterns were calculated using finite-difference time-domain (FDTD) simulations based on the principle of reciprocity~\cite{Landau1984, Zhang2015}. The simulated area included one unit cell of the array (see Methods for details). For the period of 300~nm [Fig.~\ref{fig:2}(a)], all 1st DO SLR modes lie on the short wavelength side of the emission maximum ($\sim 580$~nm), leading to enhancement in those modes and blue-shifting of the emission. We speculate that the enhancement observed near the modes TM$_0^{\pm 1, 0}$ and TM$_0^{0, \pm 1}$ may arise from the outcoupling of the SPP mode. Comparing to Fig.~\ref{fig:2}(b), it is evident that the emission pattern can be tuned by adjusting the array period: increasing the period to 370~nm red-shifts the dominant SLR modes (TE$_0^{\pm 1, 0}$ and TE$_0^{0, \pm 1}$) to overlap with the emitter peak, resulting in stronger intensity enhancement. The dispersions of the SLR modes are clearly visible in the EL. In this case, the maximum enhancement achieved is around 30~\%. Tuning the array period shifts the SLR modes and alters the emission profile, demonstrating tunability via array geometry. The enhancement or de-enhancement depends on the overlap between the SLR resonance and the emission spectrum, as well as the applied current.

The measured features correspond closely to the dispersions of the array modes associated with the first DOs. The simulations show good agreement with experiments in terms of mode location; however, some qualitative differences are present. In experimental data, the (doubly degenerate) mode TE$_0^{0, \pm 1}$ for the $p=370$~nm array is broader than simulations predict, whereas it is conversely narrower and weaker for $p=300$~nm. These discrepancies likely arise because the simulations are purely optical and do not consider any electrical modifications to the devices with the inclusion of nanoparticles. 

Figure~\ref{fig:2}(c) shows the experimental spectra of Figures~\ref{fig:2}(a) and~\ref{fig:2}(b) integrated over angles $\pm 13 \degree$ and normalized to the SL-REF device. The $p=370$~nm array (golden curve) results in enhanced emission with an overall enhancement factor (EF) of 1.06, with a maximum of 1.16 at the wavelength of 575~nm. In turn, the $p=300$~nm array (blue curve) leads to a decrease in intensity with an overall EF of 0.69 and a maximum of 0.95 at 553~nm, clearly shifted from the maximum of the reference (black dash-dotted curve) at 570~nm. 

Extended characterization of angle-resolved spectra for varying periods is presented in Figures~\ref{fig:SI-Aalto_all_dispersions}. The integrated spectrum, EF over the whole measured region, and EF from a narrower region around the maximum enhancement are then presented in Figure~\ref{fig:SI-Aalto_Integrated}. Interestingly, the $p=300$ nm array, which shows the lowest EF over the whole measured region, conversely shows the highest EF in the narrow region. This indicates that careful spectral overlap and the use of narrow emitters could significantly improve the extraction efficiency of the NAROLED devices beyond the observed values.

The nanoparticles embedded in the D2- and DQ-NAROLEDs had a height of 30~nm and a diameter of 100~nm. The array periods were 310~nm for the D2-NAROLED and 320~nm for the DQ-NAROLED. Figures~\ref{fig:2}(d) and~\ref{fig:2}(e) display the angle-resolved EL from the D2- and DQ-NAROLED devices from goniometric measurements, revealing dispersive SLR modes absent in the REF devices (shown in the insets). Again, the overlap between the emission features and the dispersive modes from the theoretical model is very good. Overall, the effect of the nanoparticle arrays is weaker in D2-NAROLED [Figure~\ref{fig:2}(d)] than in DQ-NAROLED [Figure~\ref{fig:2}(e)]. Normalized, angle-integrated spectra are shown in Figure~\ref{fig:2}(f). The D2-NAROLED exhibits only negligible changes compared to the reference device. In contrast, the DQ-NAROLED reveals a slightly increased emission across a broad wavelength range relative to the DQ-REF device. Compared to SL-NAROLED, whose peak enhancement of 0.3 is larger than the peak enhancement of DQ-NAROLED (0.15), however, along the SLR modes away from $\Gamma$-point, they are comparable. Furthermore, the
measured angular range from $\pm 75 \degree$ shows the complicated mode structure better, and that the SLRs also change the emission pattern for large angles. Further characterization of the DQ-devices with array periods of 330~nm and 340~nm is presented in Figures~\ref{fig:SI-KTU_All_periods} of the Supporting Information. 

The DD-NAROLEDs were characterized similarly to the SL-NAROLEDs using Fourier-plane imaging through a linear polarizer. The overall characteristics of the emission patterns in the DD-NAROLED are similar to those in the SL-, D2, and DQ-NAROLEDs, and will be discussed more thoroughly in the following section (see also Fig.~\ref{fig:3} and Fig.~\ref{fig:SI-UTU-measurements}(c-d)). Overall, the effect of the embedded nanoarrays on the emission directionality is similar, independent of the exact device structure. In all cases, the emission follows the SLRs, and by carefully overlapping the spectral position of the SLRs with the emission peak of the emitting material, the extraction efficiency into specific directions can be controlled.

\subsection*{Linear Polarization Tunability}
\begin{figure}[H]
\includegraphics[width=\textwidth]{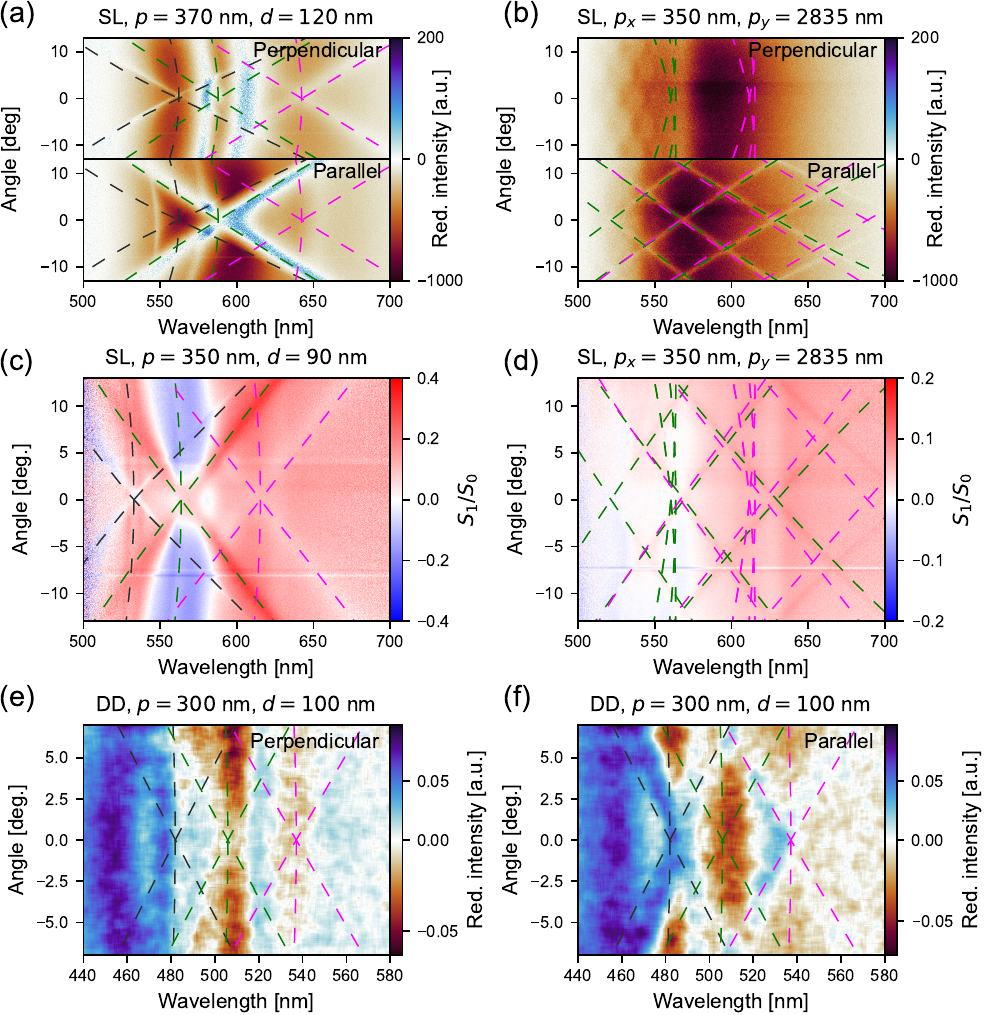}
\caption{Polarized angle-resolved reference-reduced EL. (a) SL-NAROLED with a square array at \SI{25}{\milli\ampere\per\centi\meter\squared}. (b) SL-NAROLED with a high-aspect-ratio rectangular (chain) lattice at \SI{75}{\milli\ampere\per\centi\meter\squared}. (c, d) Degree of linear polarization as the ratio of the Stokes parameters $S_1 / S_0$ of the SL-NAROLEDs with a square lattice in panel (c) and a rectangular lattice [same as in panel (b)] in (d), at \SI{50}{\milli\ampere\per\centi\meter\squared}. Perpendicular-polarized (e) and parallel-polarized (f) angle-resolved EL spectra of DD-NAROLEDs with a square array at \SI{0.25}{\milli\ampere\per\centi\meter\squared}. The dashed lines show the DOs [Eq.\eqref{e:dispersion_rectangular}]. The magenta, green, and black lines correspond to modes propagating with the effective index of the TM$_0$, TE$_0$, and TM$_1$ waveguide modes, respectively (see Methods for details on the calculation of the waveguide mode indices). The array periods $p$, $p_x$, and $p_y$ and particle diameters $d$ are indicated in panel titles. The data in (e, f) has been filtered using a Savitzky-Golay filter to illustrate the signal more clearly. Color scale for (a,b) and (e,f) in arbitrary units. }
\label{fig:3}
\end{figure}

The polarization control of the NAROLEDs was studied by measuring the angle-resolved EL spectrum through a linear polarizer. Figures~\ref{fig:3}(a--b) show the polarized emission of SL-NAROLED devices with two different geometries. In Figure~\ref{fig:3}(a), the nanostructures are arranged in a square geometry with a period of 370~nm and a particle diameter of 120~nm. Both the SL-NAROLED and SL-REF were measured at a constant current density of \SI{25}{\milli\ampere\per\centi\meter\squared}. The colorscale represents reduced intensity relative to SL-REF (shown in the insets) for both polarization components.

For both measured polarization axes, the polarizations of the TM$_{1}^{m_1,m_2}$ and the TM$_{0}^{m_1,m_2}$ are the opposite to the TE$_{0}^{m_1,m_2}$ mode. In the perpendicular polarization, the TE$_{0}^{\pm 1, 0}$ is seen to split into two. This is explained by a sample tilt about the axis of the spectrometer slit: disregarding any effects from the intermediate optics, the tilt about the slit axis results in the imaged spectra collected from $k_x=k_0\sin\theta_x$ instead of $k_x=0$. Consequently, the parabolic TM modes with DOs of ($m_1=\pm 1$, $m_2 = 0$) are no longer degenerate, while the TE modes associated with DOs (0, $\pm 1$) are almost unchanged.

Presenting a nanoparticle arrangement designed for creating angularly uniform emission, i.e., a flat band, Fig.~\ref{fig:3}b shows the dispersion of an SL-NAROLED incorporating a rectangular lattice with asymmetric periods $p_x = 350$~nm and $p_y = 2835$~nm (aspect ratio of $p_x:p_y$ = 1:8.1). The resulting chain lattice geometry is shown in Figure~\ref{fig:SI-Chainlattice Schematic}. The SL-NAROLED and SL-REF devices were measured at a constant current density of \SI{75}{\milli\ampere\per\centi\meter\squared} to improve visibility of the weaker modes. The large disparity between the periods leads to narrowing of the Brillouin zone along one reciprocal lattice direction, visible in the parallel component as replicas of the linear TE modes at angles $\sim \pm 12 \degree$. In the perpendicular component, a flat band forms as a result of superposed TM modes, which extends over all angles~\cite{Heilmann2026, Lehikoinen2026}. The flat band is of particular interest in terms of narrowing the emission spectrum, with the dispersionless nature of the band additionally providing uniform color to all angles of emission. However, the narrowed Brillouin zone comes with a trade-off: a reduced filling factor that decreases the strength of these modes. Nevertheless, this observation demonstrates the versatility of embedded nanostructures for controlling emission via geometry tuning. Lower-loss dielectric nanoparticles, where particle size can be tuned for stronger scattering, can enhance the mode strength of the flat band. 

The degree of linear polarization is quantified as the ratio of the Stokes parameters, $S_1/S_0 = (I_p-I_s) / (I_p+I_s)$, where the subscripts p and s refer to parallel and perpendicular polarization, respectively. For SL-NAROLEDs in the square and the rectangular chain lattice geometries, the degree of linear polarization is shown in Figures~\ref{fig:3}(c-d), respectively. For the SL-NAROLED in a square geometry of period p = 350 nm and particle diameter of 90~nm (measured at \SI{50}{\milli\ampere\per\centi\meter\squared}) in panel Figure~\ref{fig:3}(c), the degree of linear polarization reaches a maximum of $S_1/S_0=0.3$. Conversely, in Figure~\ref{fig:3}(d), two flat bands at the refractive indices of the first DOs of the fundamental TE$_0$ and TM$_0$ waveguide modes are seen. 

The polarization properties of the DD-NAROLED were similarly investigated using Fourier-plane imaging through a linear polarizer. Nanostructures with a diameter of 100~nm and a height of 30~nm were fabricated with periodicities of 300~nm and 320~nm. For the nanoarrays with a period of 300~nm, the $\Gamma$-point of the dominant SLR mode was well aligned with the peak emission wavelength of the blue emitter. Figures~\ref{fig:3}(e--f) show the perpendicular and parallel components of the polarized angle-resolved EL over an angular range of $\pm 11.54 \degree$. Distinct spectra are observed for the two polarizations with the hybridized TM$_{1}^{m_1,m_2}$, TE$_{0}^{m_1,m_2}$, and TM$_{0}^{m_1,m_2}$ modes, similar to the SL-NAROLED. Here, the theoretically calculated modes are obtained using the refractive indices of the waveguide modes of the DD-OLED as presented in Figure~\ref{fig:effective_indices}(b) of Methods. The DO curves in the calculated refractive indices are presented with the TM$_1$ (black dashed), TE$_0$ (green dashed), and TM$_0$ (magenta dashed) modes, respectively. Results for array parameters (a period of 320~nm) where the SLR modes are not as well aligned with the emission spectrum of the blue emitter are shown in Figure~\ref{fig:SI-UTU-measurements}. 

The degree of linear polarization of the D2- and DQ-devices was not studied in detail. However, as can be seen in Figure~\ref{fig:2}, the emission features correspond closely to the dispersion features from the theoretical model. Hence, the overall polarization characteristics are the same as in the SL- and DD-NAROLEDs. 

\subsection*{Efficiency Enhancement}

\begin{figure}[H]
\includegraphics[width=\textwidth]{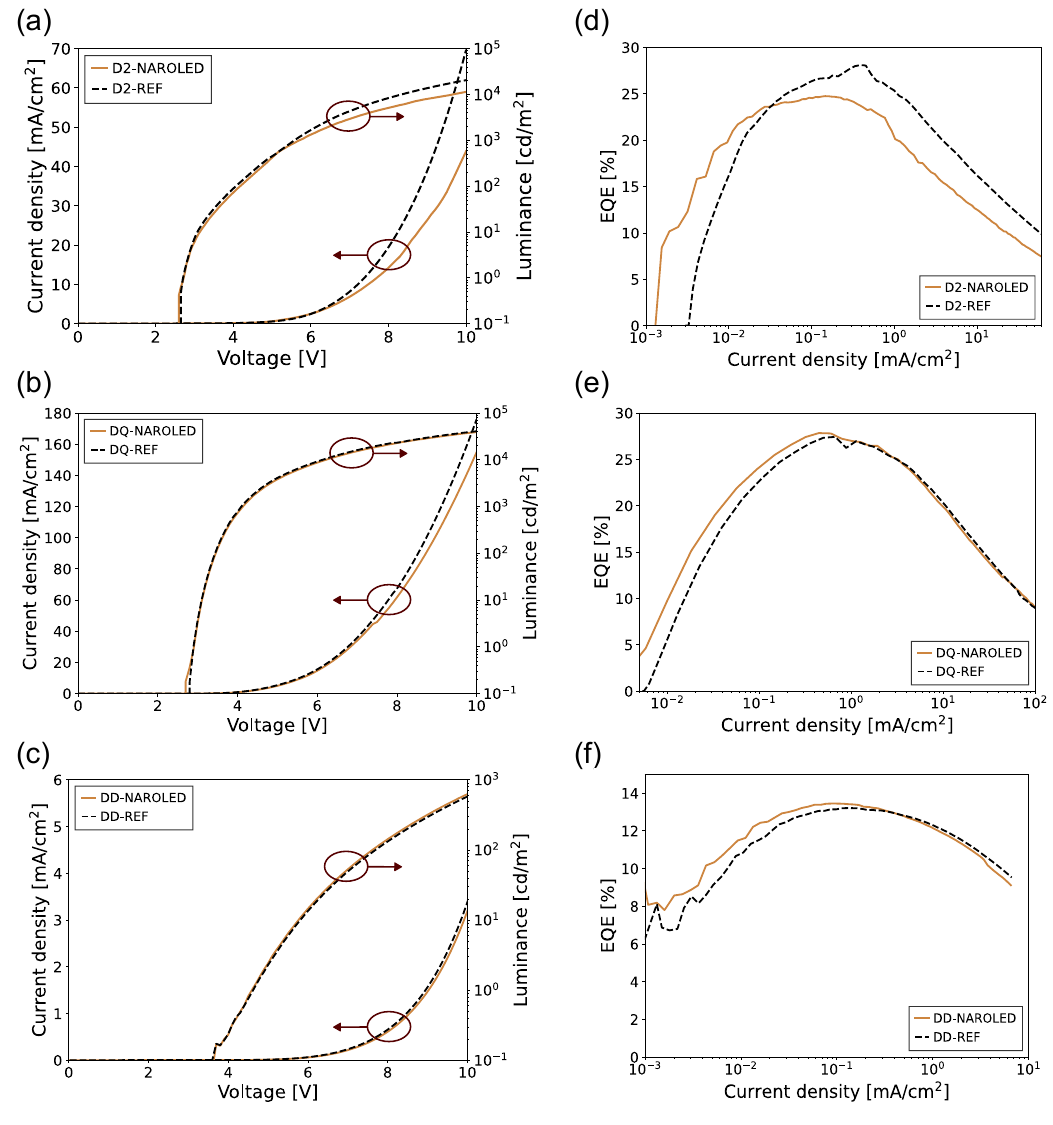}

\caption{J-V-L curves of nanostructured and reference devices for (a) D2, (b) DQ, and (c) DD devices. (d)-(f) EQE dependence on the current density for the devices of (a)-(c), respectively. 
%(c) Polar emission plots showing the photon density dependence on viewing angle for DQ devices, measured at a constant current density of \SI{2.5}{\milli\ampere\per\centi\meter\squared} (d) J-V-L and (e) EQE vs current density curves of DD in the applied voltage range from 0 to 11~V. (f) Integrated perpendicular-polarized and parallel-polarized EL spectra of the DD-NAROLED normalized to the pure reference (yellow solid curves) and the pure reference (black dashed curves), recorded under identical k-space sampling conditions: slit width of \SI{200}{\micro\meter}, applied current of \SI{10}{\micro\ampere}, and an exposure time of \SI{0.5}{\second}.
}
\label{fig:4}
\end{figure}

The performance of the D2-, DQ-, and DD-devices was studied with J-V-L characteristics in Figures~\ref{fig:4}(a-c) and efficiency enhancements with EQE dependence on the current density in Figures~\ref{fig:4}(d-f), respectively. The J-V curves of the SL-NAROLED are discussed in the section Device Stability and shown in the SI Fig.~\ref{fig:SI - JV Curves}. 

The electrical characteristics of the D2-NAROLED in Figure~\ref{fig:4}(a) closely compare to those of the D2-REF up until around 6~V, after which it begins to deviate. The luminance of D2-NAROLED follows that of the REF throughout the measurement range. Similarly, for the DQ NAROLED presented in Figure~\ref{fig:4}(b), the current density starts to deviate at higher voltages, though this happens later (around 7~V) and the effect is less drastic than for the D2-NAROLED. However, the luminance closely follows that of the DQ-REF throughout the measured region. For the DD-NAROLED presented in Figure~\ref{fig:4}(c), both the current density and the luminance are similar to the respective quantities of the DD-REF. This is presumably due to the quite low current density of the DD devices at 10~V. 

The operational differences and efficiency behaviour are more pronounced in the EQE curves. In Figure~\ref{fig:4}(d), the D2-NAROLED exhibits a large increase in efficiency (from 13.2~\% to 19.3~\%) at low current densities, followed by lower efficiency after around \SI{0.02}{\milli\ampere\per\centi\meter\squared} as compared to the D2-REF. A similar qualitative trend can be seen for both the DQ-NAROLED in Figure~\ref{fig:4}(e) (enhancement from 13.2~\% to 15.2~\%) and DD-NAROLED in Figure~\ref{fig:4}(f) (enhancement from 8.6~\% to 10.2~\%), where an increase in EQE can be observed at low current density regions. However, here, there is no de-enhancement at high current densities. 

Overall, we observe consistent J-V-L and EQE behaviour across all the studied OLED devices. The electrical characteristics of the NAROLEDs are stable and closely resemble the behaviour of the references. Also, each NAROLED exhibits EQE enhancements right after the turn-on region and slightly worse roll-off at high current densities. These measurements indicate that the electrical performance of OLEDs is not harmed and can even be improved at low current regimes by the inclusion of the embedded nanostructures. 

%This behavior indicates that while the optical outcoupling benefit provided by the arrays remains constant, it becomes negligible relative to the electrical efficiency roll-off (quenching and resistive losses) that dominates the TADF system at high brightness. These results establish that integrating plasmonic lattices is a robust strategy for recovering trapped energy, though the magnitude of the gain is fundamentally limited by the pre-existing outcoupling efficiency of the emitter orientation.

The SL-, D2-, and DQ-NAROLEDs were stable over repeated operations, e.g., the SL-NAROLED functioned well for hundreds of current sweeps. In contrast, the DD-NAROLED, which used a blue emitter, decayed fast due to the intrinsic instability of the DMAC-DPS emitter. 

\section*{Discussion and Conclusions}

We studied the tunability of OLED emission via embedded plasmonic nanoparticle arrays inside the stack in four state-of-the-art devices. The SLR modes hosted by the arrays were demonstrated to provide control over the directionality, spectral shape, and polarization of the emission. Furthermore, significant enhancement in specific modes at operational current densities, as well as EQE enhancement at low current densities, were observed.

The performance of the four different device types was overall similar, however, a few differences were observed. The SL-NAROLED showed the sharpest modes and strongest individual mode enhancement; this could be due to the close vicinity of the nanoparticles to the emissive layer. Three device types were stable over a large number of operations, while the DD-NAROLED decayed fast, which can be attributed to the typically weaker stability of blue emitters.

Extraction schemes for improved outcoupling can be classified into three categories: external, internal, and molecular engineering. External solutions, such as substrate surface roughening or the use of microlens arrays, can be used to extract substrate modes. Conversely, internal solutions and molecular engineering can be used to extract light from all loss channels. Each scheme can be used jointly. 

The relevant point of comparison for our nanoparticle approach is the internal extraction solutions, in terms of EQE and EL enhancement, spectral modification, as well as directional and polarization control. Internal extraction solutions such as low refractive index grids, nanostructuring of the OLED layers, and scattering layers between the substrate and the anode have been shown to enhance EQE by factors of 1.3-1.65~\cite{sun2008enhanced, chang2013nano, ou2014extremely, zhou2015efficiently}. While modification of the angular distribution is shown, purposeful control of emission angles, spectral shape, or polarization is not reported. We realized a maximum enhancement of 30~\% at near-normal directions around the wavelength of the array modes. EQE was shown to remain relatively unaffected by the presence of the nanoparticles. At low drive currents, an EQE enhancement factor of 1.46 (from 13.2~\% to 19.3~\%) was obtained. 

Spectral modification of a bottom-emitting OLED utilizing LSPRs of Ag nanodots showed EL peak red-shifting of 38 nm~\cite{lee2016effect}. LSPRs and SLRs of Al nanodiscs have additionally been shown to blue-shift emission from a bottom-emitting OLED~\cite{Auer-Berger2017-2}. However, LSPRs do not offer directional control, and increasing the dot size lowered power efficiency.  In our work, spectral modification by array geometry tuning was seen with all studied devices, for example, observation of angularly uniform flat band emission from an asymmetric chain lattice, and a peak position shift from 507~nm to 522~nm with a square geometry.

Polarization control with corrugated OLEDs has been reported to achieve a degree of linear polarization of 0.95 in bottom-emitting OLEDs~\cite{chen2024dual}. The resulting emission is, by necessity, very dispersive due to corrugation. We observed polarized response with a degree of linear polarization, \textit{i.e.}, an $S_1$/$S_0$ ratio, of 0.3 at the strongest modes.

As indicated, nanoparticle arrays can provide an internal solution capable of controlling all the previously mentioned parameters while also improving the efficiency of OLEDs. SLRs from periodic Al nanodisk arrays have been shown to tune the color coordinates and improve the current efficiency by 23~\% in white OLED devices~\cite{Auer-Berger2022}. However, the EQE and polarization control were not quantified, and the array design with respect to mode engineering was not systematically studied.

Here, we established a framework for array design principles based on OLED waveguide modes, emitter spectrum, and array geometry. We show that the design principles are general across four different high-performing OLEDs. We found that emission enhancement is generally strongest when the array parameters are chosen such that there are SLR modes at the energies near the emission maximum of the emitter. On the other hand, de-enhancement may be seen at other energies. A promising next step is thus to embed the nanoparticle arrays in narrow emitter OLEDs, with improved mode overlap, to obtain maximal efficiency enhancements. Another viable prospect is the use of lower-loss dielectric nanoparticles to avoid ohmic losses. The excellent match of the experiments with simulation and theory provides the basis for rational design of the desired emission properties. The combination of the empty lattice approximation with the waveguide mode indices was shown to explain the main observed features of the emission: this gives an intuitive, fast, and computationally light tool for qualitative design of the required array geometry. These lightweight simulations thus provide the means to adapt nanoparticle array geometries to other OLED designs. Quantitative effects such as the degree of polarization or amount of enhancement can then be tested by FDTD and reciprocity simulations, which were shown to have fair agreement with the experiment as well. 

In summary, compared to the numbers presented in the previous paragraphs, the performance of our devices is typically similar, although occasionally slightly better or worse, in \textit{all} the features considered. The reported values in our studies were obtained from trivial square array geometries. Despite the unoptimized array geometry, we saw improvements in all of the measured quantities. Thus, our results verify embedded nanoparticles and mode engineering as a viable framework for internal efficiency enhancement and emission control of OLEDs. In general, the array geometry, particle shape, and material can be tuned for optimized control of a specific quantity. Nanoparticle arrays offer special opportunities for independently tailoring the direction, color, and polarization of the emission, exemplified by the new flat band chain lattice concept~\cite{Lehikoinen2026,Heilmann2026}, which we successfully embedded in an OLED device here. 

The research towards increasing OLED efficiency focuses considerably on molecular engineering of the devices and on various external components. The high-end materials remain expensive to synthesize, and the external components add complexity to the fabrication as well as incompatibility with the demanding form factor requirements. While the nanoparticle arrays were produced here via an EBL process, in mass manufacturing, nano-imprint lithography would present a direct plug-in step to the existing fabrication flow. Moreover, the embedded nanostructures are not device-specific. They could be implemented in existing OLED devices via the previously explained mode engineering and combined with other existing solutions. Nanoparticle arrays and their SLR modes have been widely explored when combined with optically pumped gain medium, leading to lasing and Bose-Einstein condensation~\cite{zhou2013lasing,wang2018rich,freire2025plasmonic,hakala2018bose}, however, they have scarcely been utilized in organic optoelectronic devices. Our results highlight the feasibility and prospects of this research direction. 

\section*{Methods}

\subsection*{Nanostructure fabrication}
Each device went through an identical cleaning procedure consisting of subsequent ultrasonications in 2~\% Hellmanex in heated DIW (10~min), dump-rinse of hot DIW repeated twice, heated DIW (5~min) repeated three times with fresh DIW, acetone (10~min), IPA (15~minutes), and lastly, dried with N$_2$. 

Nanofabrication was performed immediately thereafter, starting with spin-coating of PMMA A4 at 3000 RPM for 60 seconds (static dispense). Before EBL exposure, a 10-nm aluminum film was deposited on the resist by E-beam evaporation. This layer acted as a charge-dissipation layer, improving pattern accuracy and reducing damage to the ITO layer from accumulated charges. The patterning was then performed with a Raith EBPG5200 EBL system with 100 kV acceleration voltage, 5 nA beam current, 300~nm beam aperture, and \SI{1500}{\micro\coulomb} dose. After patterning, the aluminum film was etched off in a diluted 30~\% AZ351B:DIW solution for 2 minutes, and the resist was developed in a 1:3 MIBK:IPA solution for 30 seconds. The development was stopped with an immersion in IPA for $\geq30$ seconds. The aluminum nanoparticles were then deposited in an E-beam evaporator to a thickness of 30 nm. Here, an additional adhesion layer of commonly used materials (Ti/Cr) was not necessary, as aluminum exhibited strong adhesion to ITO on its own. Lastly, lift-off of excess material was performed in an acetone soak, either for a few hours in warm (45$\degree$C) acetone or overnight in room-temperature acetone.

The quality and homogeneity of the sample fabrication were measured with a commercial scatterometry tool (LabScatter, TeraNova), confirming excellent array homogeneity for the entire area, with a retrieved particle diameter of $124.2\pm 0.2$~nm and particle height of $30.2\pm 0.9$~nm (see Supplementary Information for details).

\subsection*{OLED fabrication and specifications}
\subsubsection*{SL-OLED}
Prior to the OLED stack evaporation, all SL-OLED devices went through O2-plasma treatment at a plasma power of 100 W for 1-2 minutes. This had a two-fold effect: cleaning the sample of possible remaining organic residue from previous steps as well as increasing the wettability of the sample surface for spin-coating of PEDOT:PSS (AI 4083 from Ossila). Prior to spin-coating, the PEDOT:PSS was filtered through a \SI{0.45}{\micro\meter} PES filter. Spin parameters of 3000 RPM for 60 seconds were used, yielding a layer thickness of 40~nm. Then, the edges of the substrate were wiped with DIW-soaked cleanroom swabs to expose the electrode strips, while taking care that the active area remained intact. Lastly, the film was fully dried by placing it on a hot plate at 150$\celsius$ for 5 minutes. Up until this point, all processing steps were done in ambient conditions. The subsequent layers were then deposited in a physical vapor deposition (PVD) system, at a base pressure of 3--6$\times10^{-7}$~mbar, using material-specific recipes. Active layers and cathode strips were deposited through their respective shadow masks.

The SL device structure consists of 100~nm pre-patterned ITO as the transparent anode; a 40~nm layer of poly(3,4-ethylenedioxythiophene):polystyrene sulfonate (PEDOT:PSS) as a smoothing layer and hole injection layer (HIL); a 6~nm layer of molybdenum trioxide (MoO$_3$) and a 3~nm layer of buckminsterfullerene (C60) as an anode-side interlayer (IL); a 75~nm layer of 5,10-Bis(4-(9H-carbazol-9-yl)-2,6-dimethylphenyl)-5,10-dihydroboranthrene (CzDBA) as the emissive layer (EML); 4~nm layer of 2,2',2"-(1,3,5-Benzinetriyl)-tris(1-phenyl-1-H-benzimidazole) (TPBi) as the cathode-side IL; and finally the device was capped off with a 100~nm thick aluminum layer as the cathode. The pre-patterned substrate and the layer materials (sublimed-grade) were all purchased from Ossila.

\subsubsection*{DD-OLED}
The nanopatterned ITO substrates were subjected to UV–ozone treatment (UVC-1014) for 15 minutes in order to remove organic residues and enhance surface cleanliness prior to the fabrication of DD–OLED devices. The PEDOT:PSS layers of DD-OLED devices were prepared using the same recipe and processing parameters, such as solution filtration, spin-coating parameters, and thermal annealing, as the SL-OLED devices. An Angstrom thermal evaporation system connected to a glovebox with a fully automated mask exchange was used to deposit all OLED layers at a base pressure of $2 \times 10^{-7}\,\mathrm{mbar}$.
The device structure of the DD-OLED consists of ITO as the transparent anode; a 40~nm layer of poly(3,4-ethylenedioxythiophene):polystyrene sulfonate (PEDOT:PSS) as a smoothing layer and HIL; a 5~nm layer of molybdenum trioxide (MoO$_3$) as an additional HIL; a 40~nm layer of 1,3-bis(N-carbazolyl)benzene (mCP) serving as the hole transport layer (HTL); a 65~nm neat film of 10,10$^\prime$-(4,4$^\prime$-sulfonylbis(4,1-phenylene))bis(9,9-dimethyl-9,10-dihydroacridine) (DMAC–DPS) as the blue emitter; a 50~nm layer of bis[2-(diphenylphos\-phino)phenyl] ether oxide (DPEPO) as the electron transport layer (ETL); a 1~nm layer of lithium fluoride (LiF) as the electron injection layer (EIL); and finally a 110~nm aluminum cathode.

\subsubsection*{D2- and DQ-OLEDs}
DACT-II and DQBC-based multilayer devices were prepared using a hybrid deposition approach. Organic semiconductor materials, including the TADF emitters 9-[4-(4,6-diphenyl-1,3,5-triazin-2-yl)phenyl]-N,N,N$'$,N$'$-tetraphenyl-9H-carbazole-3,6-diamine (DACT-II) and\linebreak DQBC 3,11-di(10H-phenoxazin-10-yl)dibenzo[a,c]phenazine (DQBC), were purchased from commercial suppliers (Lumtec, OSSILA, MERCK, sublimed grade) and used as received without further purification.  The patterned substrates were washed with the different solvents (acetone, isopropanol), followed by UV-ozone treatment. A 40~nm thick HIL of PEDOT:PSS was spin-coated onto the substrates and annealed at 140 °C for 15 minutes to remove solvent and planarize the surface.
The subsequent organic layers were deposited via thermal evaporation in a vacuum chamber at a base pressure of 2 × $10^{-6}$ mbar. The device stack consisted of a 40~nm (HTL) of N,N$'$-di(1-naphthyl)-N,N$'$-diphenyl-(1,1$'$-biphenyl)-4,4$'$-diamine (NPB) and a 4~nm exciton-blocking layer of 4,4$'$,4"-tris(carbazol-9-yl)triphenylamine (TCTA). A 30 nm EML was formed by co-depositing either DACT-II doped at 10~wt\% into a host of 2,3,4,5,6-pentakis(3,6-di-tert-butyl-9H-carbazol-9-yl)benzonitrile (5tCzBN), or DQBC doped at 10~wt\% into 3,6-bis(carbazol-9-yl)-9-(2-ethyl-hexyl)-9H-carbazole (TCz1). The active stack was completed with a 4~nm hole-blocking layer of diphenyl[4-(triphenylsil\-yl)phenyl]phosphine oxide (TSPO1) and a 40~nm ETL of 2,2$'$,2$''$-(1,3,5-benzinetriyl)-tris(1-phenyl-1-H-benzimidazole) (TPBi). The cathode was formed by a 1.5~nm layer of lithium fluoride (LiF) and 100~nm of aluminum (Al).
 
\subsection*{Characterization of OLEDs}
The SL devices presented in Figure \ref{fig:1}b were characterized in ambient conditions. The single-layer architecture with CzDBA as the emissive layer has good stability, and no signs of degradation were observed, even after prolonged active operation of 50 repeated JV-sweep cycles from 0-8V without a cooling time between sweeps. The current-voltage characteristics were recorded with a Keithley 2600B source-meter by utilizing a stepped-continuous voltage supply mode. The angle-resolved EL spectra were captured with an in-house built Fourier-plane imaging setup, where light was collected using a 0.3 NA objective. The back-focal plane of the objective was then transferred via lenses to a spectrometer (Princeton Instruments), where a vertical slit in front of it determines the direction in which different angular contributions of the emission are collected, corresponding to a specific component of the in-plane wavevector in the sample (in our notation, $k_y$). Here, the OLED pixels were activated with constant current throughout each individual measurement, and measurements were repeated for a range of different currents.

The DDs were characterized inside a glovebox using an in-house engineered measurement setup equipped with a Keysight B2902B source meter for recording current–voltage data. Since blue TADF OLEDs are prone to relatively rapid degradation under electrical bias~\cite{Tankeleviciute2024,Lee2019}, we employed a two-regime J–V–L measurement protocol. In the first regime (–2 to 3 V), the J–V sweep was performed using 1-second electrical pulses to allow the source-meter to accurately measure the very low current densities when the OLED is off or near turn-on. In the second regime (3 to 5 V), where the OLED is emissive, the pulse width was reduced to 200 ms to minimize device degradation during measurement. Photocurrent was measured using a calibrated large-area silicon photodiode 10 × 10 mm². The photodiode was positioned approximately 0.1 mm above the emitting pixel to collect forward-emitted photons. EL spectra were recorded at normal incidence (0°) with an Ocean Optics USB2000+ spectrometer. Using the measured photocurrent together with the 0° EL spectra and the device current, we calculated the luminances and efficiencies of fabricated OLEDs.

Current density-voltage and luminance-voltage characteristics of D2- and DQ-NAROLEDs were recorded using a calibrated silicon photodiode (PH100-Si-HA-D0) coupled with a Keithley 2400C source meter. EL spectra were acquired using an Avantes AvaSpec-2048XL spectrometer. Angular dependent EL (ADEL) measurements were performed using a Fluxim Phelos goniometer system.  Device efficiencies were estimated from the luminance, current density, and EL spectrum. 

\subsection*{Thin-film optical characterization}

Thin-film thicknesses and the complex refractive index data of the SL-devices were measured with SE-2000 Spectroscopic ellipsometer in the spectral range of 250--885~nm. Thin film thicknesses and optical constants of the DD-device materials were measured using a J.A. Woollam VASE ellipsometer equipped with a xenon lamp, covering a spectral range of 250--2500~nm.

\subsection*{Electromagnetic field simulations}

The finite-difference time-domain (FDTD) simulations were implemented in Lumerical 3D Electromagnetic Simulator (Lumerical Inc.), version 2025-R1.1. The far field emission pattern was computed using the principle of reciprocity~\cite{Landau1984, Zhang2015} and BFAST~\cite{Liang2014} to ensure constant-angle injection. The simulated area included one unit cell of the nanoparticle lattice, with BFAST periodic boundary conditions applied on the lateral boundaries, and metal (perfect electrical conductor) and perfectly-matched layer (PML) boundary conditions on the aluminum and glass boundaries, respectively. To account for the photoluminescence spectrum of the emitter, the computed emission intensity as a function of wavelength was weighted by the measured photoluminescence spectrum of CzDBA. The following sources for material data (refractive index) were used: aluminum~\cite{Palik}, CzDBA~\cite{Li2021}, C$_{60}$~\cite{Sittinger2022}, MoO$_3$~\cite{Vos2015}, PEDOT:PSS~\cite{Chen2015}, ITO~\cite{Konig2014}. Refractive index data of TPBi was measured using ellipsometry. For the substrate, a constant index of 1.52 was used. The mesh element size was (2.5~nm)$^3$ in the EML, with grading applied in the $z$-direction (the plane of the array was $x$-$y$).

\subsection*{Surface lattice resonance dispersions and waveguide modes}

Surface lattice resonances (SLRs) arise from the coupling (hybridisation) of localized surface plasmon resonances (LSPRs) of individual nanoparticles with in-plane (grazing incidence) DOs of a periodic lattice. Within the empty lattice approximation, the dispersion of the SLR mode related to the diffraction order $(m_1, m_2) \in \mathbb{Z}^2$ of a rectangular array with periods $p_x$ and $p_y$ is
\begin{equation}
    \label{e:dispersion_rectangular}
    E(\mathbf{k}_\parallel) = \frac{\hbar c_0}{n} \sqrt{\left(k_x + m_1 \frac{2\pi}{p_x}\right)^2 + \left(k_y + m_2 \frac{2\pi}{p_y}\right)^2},    
\end{equation}
where $\mathbf{k}_\parallel = (k_x, k_y)^\text{T}$, $\hbar$ is the reduced Planck's constant, $c_0$ the speed of light in vacuum, and $n$ the refractive index in which the mode propagates~\cite{guo2017geometry}. From this, it is evident that for a square array ($p_x = p_y \equiv p$), at the $\Gamma$ point, the first DOs $(\pm 1, 0)$ and $(0, \pm 1)$ are degenerate at the wavelength
\begin{equation}
    \label{e:gamma_point}
    \lambda = n p.    
\end{equation}
Per Eq.~\eqref{e:dispersion_rectangular}, diffraction in the plane of incidence yields a linear dispersion: for example, if the plane of incidence is in the direction of $k_y$, diffraction in that plane means $m_2 \neq 0$, $m_1=0$, and $k_x=0$, leading to a linear dependence of the energy on $k_y$. The resulting pair of crossing lines is called TE modes. On the other hand, diffraction out of the plane of incidence gives a doubly degenerate, curved dispersion, called a TM mode. 

The multi-layered OLED structures form a waveguide that supports guided modes whose effective indices $n_\text{eff}(\lambda)$ depend on the composition of the OLED stack~\cite{Adams1981}; the effective indices are calculated as described below. The SLRs and waveguided modes may hybridize into quasi-guided modes~\cite{Murai2013}. Quasi-guided modes can be identified from the experimental data by associating the observed modes to the dispersions of Eq.~\eqref{e:dispersion_rectangular}, where the effective index of the associated waveguided mode $n_\text{eff}$ is used as the refractive index in Eq.~\eqref{e:gamma_point}. Coupling to the different waveguide modes explains the observed multiplication of array modes, see e.g. Fig.~\ref{fig:2}(a) and~(b). Often in metal-clad waveguides, like OLEDs, the fundamental TM$_0$ waveguide mode is hybridized with the SPP mode at the electrode--organic interface~\cite{Adams1981}. Via that hybridization, diffraction by the nanoparticle array may help outcouple the SPP mode.

\begin{figure}[H]
    \centering
    \includegraphics[width=\linewidth]{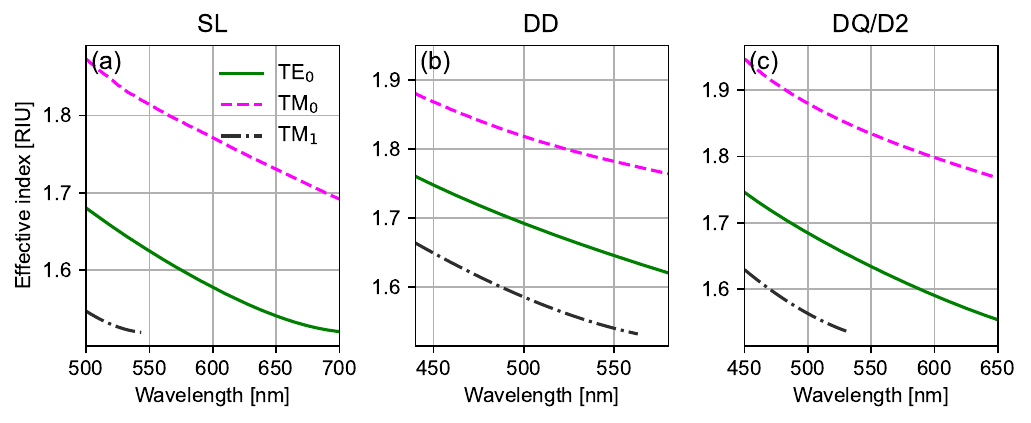}
    \caption{Effective indices of the waveguide modes of the investigated OLED structures. }
    \label{fig:effective_indices}
\end{figure}

The effective indices of the waveguide modes of the investigated OLED structures (Fig.~\ref{fig:effective_indices}) were found using finite element method (FEM) simulations implemented in COMSOL Multiphysics\textregistered~\cite{Comsol}, following the procedure outlined in Refs.~\cite{ComsolModeAnalysis, ComsolModeTracking}. 

\section*{Data and Code Availability}

The data underlying this study are openly available in Zenodo~\cite{Dahlberg2026Enhancement}. 

\section*{Acknowledgements}

Funded by the European Union. Views and opinions expressed are, however, those of the author(s) only and do not necessarily reflect those of the European Union or the European Innovation Council and the SMEs Executive Agency (EISMEA). Neither the European Union nor the granting authority can be held responsible for them (SCOLED, Grant Agreement No. 101098813).

The work was also supported by the Jane and Aatos Erkko Foundation and the Technology Industries of Finland Centennial Foundation as part of the Future Makers funding program, by the Research Council of Finland under project number 349313, and by the Research Council of Finland through the Finnish Quantum Flagship project 358877. The work is part of the Research Council of Finland Flagship Programme, Photonics Research and Innovation (PREIN), decision number 346529, Aalto University. This work is part of the Finnish Centre of Excellence in Quantum Materials (QMAT). 

K.S.D. acknowledges funding from the European Research Council through the European Union’s Horizon 2020 research and innovation program (Grant Agreement Number 948260) and from the Research Council of Finland (project grant No. 369819). Kaunas University of Technology acknowledges financial support from the Research Council of Lithuania (LMTLT) and the Ministry of Education, Science and Sport of the Republic of Lithuania, agreement No: S-A-UEI-23-1

Part of the research was performed at the OtaNano Nanofab cleanroom (Micronova Nanofabrication Centre), supported by Aalto University. The calculations presented above were performed using computer resources within the Aalto University School of Science “Science-IT” project. A.J.J.D. thanks Caterina Soldano and Matti Hokkanen for their help related to the OLED fabrication.

\section*{Author Contributions}
P.T., K.S.D., J.G.R., and W.L.B. conceived the project. P.T., K.S.D., D.V., and J.V.G. supervised the project. A.J.J.D. fabricated the nanoparticle arrays for all OLEDs. A.J.J.D. fabricated and measured the SL-NAROLED, and M.G. the D2- and DQ-NAROLED. M.M. and M.K. fabricated the DD-NAROLED, and M.M., M.K., H.A.Q., M.A.P., H.L., and O.S. measured it. J.L. made the outcoupling simulations. A.M.B. made the scatterometry measurements. E.A.M. helped with building the measurement setups for the SL-NAROLED. R.H. guided the nanoparticle array design and fabrication. A.J.J.D., J.L., P.T.,  D.V., and R.K.  wrote the manuscript, with contributions from other authors. All authors discussed the results and reviewed the manuscript. 

\section*{Competing Interests}
A.J.J.D., R. H., J. L., and P. T. declare that they are inventors in a filed PCT patent application (PCT/FI2026/050087) related to the flat band nanoparticle array concept used in this manuscript. The remaining authors declare no competing financial interests.
% State any conflicts of interest or “The authors declare no competing interests.”

\bibliography{References}

@article{kotadiya2019efficient,
  title={Efficient and Stable Single-Layer Organic Light-Emitting Diodes Based on Thermally Activated Delayed Fluorescence},
  author={Kotadiya, Naresh B and Blom, Paul WM and Wetzelaer, Gert-Jan AH},
  journal={Nature Photonics},
  volume={13},
  number={11},
  pages={765--769},
  year={2019},
  publisher={Nature Publishing Group UK London}
}

@article{Kaji2015-ic,
  title     = "Purely Organic Electroluminescent Material Realizing 100\%
               Conversion from Electricity to Light",
  author    = "Kaji, Hironori and Suzuki, Hajime and Fukushima, Tatsuya and
               Shizu, Katsuyuki and Suzuki, Katsuaki and Kubo, Shosei and
               Komino, Takeshi and Oiwa, Hajime and Suzuki, Furitsu and
               Wakamiya, Atsushi and Murata, Yasujiro and Adachi, Chihaya",
  journal   = "Nat. Commun.",
  publisher = "Springer Science and Business Media LLC",
  volume    =  6,
  number    =  1,
  pages     = "8476",
  month     =  oct,
  year      =  2015,
  copyright = "https://creativecommons.org/licenses/by/4.0",
  language  = "en"
}

@article{https://doi.org/10.1002/adma.202103293,
author = {Chen, Yang and Zhang, Dongdong and Zhang, Yuewei and Zeng, Xuan and Huang, Tianyu and Liu, Ziyang and Li, Guomeng and Duan, Lian},
title = {Approaching Nearly 40\% External Quantum Efficiency in Organic Light Emitting Diodes Utilizing a Green Thermally Activated Delayed Fluorescence Emitter with an Extended Linear Donor–Acceptor–Donor Structure},
journal = {Advanced Materials},
volume = {33},
number = {44},
pages = {2103293},
doi = {https://doi.org/10.1002/adma.202103293},
url = {https://advanced.onlinelibrary.wiley.com/doi/abs/10.1002/adma.202103293},
year = {2021}
}

@article{zhang2015nearly,
  title={Nearly 100\% Internal Quantum Efficiency in Undoped Electroluminescent Devices Employing Pure Organic Emitters.},
  author={Zhang, Qisheng and Tsang, Daniel and Kuwabara, Hirokazu and Hatae, Yasuhiro and Li, Bo and Takahashi, Takehiro and Lee, Sae Youn and Yasuda, Takuma and Adachi, Chihaya},
  journal={Advanced Materials (Deerfield Beach, Fla.)},
  volume={27},
  number={12},
  pages={2096--2100},
  year={2015}
}

@article{Zhang2014,
   author   = {Qisheng Zhang and Bo Li and Shuping Huang and Hiroko Nomura and Hiroyuki Tanaka and Chihaya Adachi},
   doi      = {10.1038/nphoton.2014.12},
   issn     = {1749-4893},
   journal  = {Nature Photonics},
   month    = {4},
   pages    = {326--332},
   publisher= {Nature Publishing Group},
   title    = {Efficient Blue Organic Light-Emitting Diodes Employing Thermally Activated Delayed Fluorescence},
   volume   = {8},
   number   = {4},
   url      = {https://doi.org/10.1038/nphoton.2014.12},
   year     = {2014}
}

@article{Kumar2025,
   author   = {Manish Kumar and Arpan Dutta and Hassan A. Qureshi and Michael A. Papachatzakis and Ahmed Gaber Abdelmagid and Konstantinos S. Daskalakis},
   doi      = {10.1002/adom.202501358},
   journal  = {Advanced Optical Materials},
   title    = {Single-Emitter White OLEDs via Microcavity Spectral Engineering},
   volume   = {13},
   number   = {28},
   pages    = {e01358},
   publisher= {Wiley},
   month    = {10},
   year     = {2025},
   url      = {https://doi.org/10.1002/adom.202501358}
}

@article{Auer-Berger2017-1,
   author = {Manuel Auer-Berger and Veronika Tretnak and Franz-Peter Wenzl and Joachim R. Krenn and Emil J. W. List-Kratochvil},
   doi = {10.1063/1.4998802},
   issn = {0003-6951},
   journal = {Applied Physics Letters},
   month = {10},
   pages = {173301},
   publisher = {American Institute of Physics Inc.},
   title = {Aluminum-Nanodisc-Induced Collective Lattice Resonances: Controlling the Light Extraction in Organic Light Emitting Diodes},
   volume = {111},
   url = {https://pubs.aip.org/apl/article/111/17/173301/1021215/Aluminum-nanodisc-induced-collective-lattice},
   year = {2017}
}

@article{Auer-Berger2017-2,
   author = {Manuel Auer-Berger and Veronika Tretnak and Franz-Peter Wenzl and Joachim R. Krenn},
   doi = {10.1117/1.OE.56.9.097102},
   issn = {0091-3286},
   journal = {Optical Engineering},
   month = {9},
   pages = {1},
   publisher = {SPIE-Intl Soc Optical Eng},
   title = {Adjusting the Emission Color of Organic Light-Emitting Diodes Through Aluminum Nanodisc Arrays},
   volume = {56},
   url = {https://www.spiedigitallibrary.org/journals/optical-engineering/volume-56/issue-09/097102/Adjusting-the-emission-color-of-organic-light-emitting-diodes-through/10.1117/1.OE.56.9.097102.full},
   year = {2017}
}

@article{Auer-Berger2022,
   author = {Manuel Auer-Berger and Veronika Tretnak and Christian Sommer and Franz-Peter Wenzl and Joachim R. Krenn and Emil J. W. List-Kratochvil},
   doi = {10.1007/s00339-022-05854-w},
   issn = {0947-8396},
   journal = {Applied Physics A},
   month = {9},
   pages = {751},
   publisher = {Springer Science and Business Media Deutschland GmbH},
   title = {Multipitched Plasmonic Nanoparticle Grating for Broadband Light Enhancement in White Light-Emitting Organic Diodes},
   volume = {128},
   url = {https://link.springer.com/10.1007/s00339-022-05854-w},
   year = {2022}
}

@article{Kang2024,
   author = {Kyungnam Kang and Inseop Byeon and Young Gu Kim and Jong‐ryul Choi and Donghyun Kim},
   doi = {10.1002/lpor.202400547},
   issn = {1863-8880},
   journal = {Laser \& Photonics Reviews},
   month = {8},
   publisher = {John Wiley and Sons Inc},
   title = {Nanostructures in Organic Light‐Emitting Diodes: Principles and Recent Advances in the Light Extraction Strategy},
   volume = {18},
   url = {https://onlinelibrary.wiley.com/doi/10.1002/lpor.202400547},
   year = {2024}
}

@article{Gather2015,
   author = {Malte C. Gather and Sebastian Reineke},
   doi = {10.1117/1.JPE.5.057607},
   issn = {1947-7988},
   journal = {Journal of Photonics for Energy},
   month = {5},
   pages = {057607},
   title = {Recent Advances in Light Outcoupling from White Organic Light-Emitting Diodes},
   volume = {5},
   url = {http://photonicsforenergy.spiedigitallibrary.org/article.aspx?doi=10.1117/1.JPE.5.057607},
   year = {2015}
}

@article{Moller2002,
   author = {S. Möller and S. R. Forrest},
   doi = {10.1063/1.1435422},
   issn = {0021-8979},
   journal = {Journal of Applied Physics},
   month = {3},
   pages = {3324-3327},
   title = {Improved Light Out-Coupling in Organic Light Emitting Diodes Employing Ordered Microlens Arrays},
   volume = {91},
   url = {https://pubs.aip.org/jap/article/91/5/3324/1032126/Improved-light-out-coupling-in-organic-light},
   year = {2002}
}

@article{Sun2008,
   author = {Yiru Sun and Stephen R. Forrest},
   doi = {10.1038/nphoton.2008.132},
   issn = {1749-4885},
   journal = {Nature Photonics},
   month = {8},
   pages = {483-487},
   title = {Enhanced Light Out-Coupling of Organic Light-Emitting Devices Using Embedded Low-Index Grids},
   volume = {2},
   url = {https://www.nature.com/articles/nphoton.2008.132},
   year = {2008}
}

@article{Zhou2011,
   author = {Junhong Zhou and Na Ai and Lei Wang and Hua Zheng and Chan Luo and Zhixiong Jiang and Shufu Yu and Yong Cao and Jian Wang},
   doi = {10.1016/j.orgel.2011.01.018},
   issn = {15661199},
   journal = {Organic Electronics},
   month = {4},
   pages = {648-653},
   publisher = {Elsevier B.V.},
   title = {Roughening the White OLED Substrate’S Surface Through Sandblasting to Improve the External Quantum Efficiency},
   volume = {12},
   url = {https://linkinghub.elsevier.com/retrieve/pii/S1566119911000334},
   year = {2011}
}

@article{Riedel2010,
   author = {Boris Riedel and Inga Kaiser and Julian Hauss and Uli Lemmer and Martina Gerken},
   doi = {10.1364/OE.18.00A631},
   issn = {1094-4087},
   journal = {Optics Express},
   month = {11},
   pages = {A631},
   title = {Improving the Outcoupling Efficiency of Indium-Tin-Oxide-Free Organic Light-Emitting Diodes via Rough Internal Interfaces},
   volume = {18},
   url = {https://opg.optica.org/oe/abstract.cfm?uri=oe-18-S4-A631},
   year = {2010}
}

@inproceedings{Gaertner2008,
   author = {Georg Gaertner and Horst Greiner},
   doi = {10.1117/12.780249},
   editor = {Paul L. Heremans and Michele Muccini and Eric A. Meulenkamp},
   booktitle = {Organic Optoelectronics and Photonics Iii},
   month = {4},
   pages = {69992T},
   publisher = {SPIE},
   title = {Light Extraction from OLEDs with (High) Index Matched Glass Substrates},
   volume = {6999},
   url = {http://proceedings.spiedigitallibrary.org/proceeding.aspx?doi=10.1117/12.780249},
   year = {2008}
}

@article{Reineke2009,
   author = {Sebastian Reineke and Frank Lindner and Gregor Schwartz and Nico Seidler and Karsten Walzer and Björn Lüssem and Karl Leo},
   doi = {10.1038/nature08003},
   issn = {0028-0836},
   issue = {7244},
   journal = {Nature},
   month = {5},
   pages = {234-238},
   pmid = {19444212},
   title = {White organic light-emitting diodes with fluorescent tube efficiency},
   volume = {459},
   url = {https://www.nature.com/articles/nature08003},
   year = {2009}
}

@article{Lupton2000,
   author = {John M. Lupton and Benjamin J. Matterson and Ifor D. W. Samuel and Michael J. Jory and William L. Barnes},
   doi = {10.1063/1.1320023},
   issn = {0003-6951},
   journal = {Applied Physics Letters},
   month = {11},
   pages = {3340-3342},
   publisher = {American Institute of Physics Inc.},
   title = {Bragg Scattering from Periodically Microstructured Light Emitting Diodes},
   volume = {77},
   url = {https://pubs.aip.org/apl/article/77/21/3340/515571/Bragg-scattering-from-periodically-microstructured},
   year = {2000}
}

@article{Youn2015,
   author = {Wooram Youn and Jinhyung Lee and Minfei Xu and Rajiv Singh and Franky So},
   doi = {10.1021/acsami.5b01533},
   issn = {1944-8244},
   journal = {ACS Applied Materials \& Interfaces},
   month = {5},
   pages = {8974-8978},
   publisher = {American Chemical Society},
   title = {Corrugated Sapphire Substrates for Organic Light-Emitting Diode Light Extraction},
   volume = {7},
   url = {https://pubs.acs.org/doi/10.1021/acsami.5b01533},
   year = {2015}
}

@article{Zeng2024,
   author = {Xin-Yi Zeng and Hong-Yi Hou and Yan-Qing Li and Jian-Xin Tang},
   doi = {10.1063/5.0201680},
   issn = {1931-9401},
   journal = {Applied Physics Reviews},
   month = {6},
   publisher = {American Institute of Physics},
   title = {Recent Progress of Metasurfaces in Light-Emitting Diodes},
   volume = {11},
   url = {https://pubs.aip.org/apr/article/11/2/021339/3299158/Recent-progress-of-metasurfaces-in-light-emitting},
   year = {2024}
}

@article{Kim2018,
   author = {Jongchan Kim and Yue Qu and Caleb Coburn and Stephen R. Forrest},
   doi = {10.1021/acsphotonics.8b00539},
   issn = {2330-4022},
   journal = {ACS Photonics},
   month = {8},
   pages = {3315-3321},
   publisher = {American Chemical Society},
   title = {Efficient Outcoupling of Organic Light-Emitting Devices Using a Light-Scattering Dielectric Layer},
   volume = {5},
   url = {https://pubs.acs.org/doi/10.1021/acsphotonics.8b00539},
   year = {2018}
}

@article{Song2018,
   author = {Jinouk Song and Kwon-Hyeon Kim and Eunhye Kim and Chang-Ki Moon and Yun-Hi Kim and Jang-Joo Kim and Seunghyup Yoo},
   doi = {10.1038/s41467-018-05671-x},
   issn = {2041-1723},
   journal = {Nature Communications},
   month = {8},
   pages = {3207},
   publisher = {Nature Publishing Group},
   title = {Lensfree OLEDs with Over 50\% External Quantum Efficiency via External Scattering and Horizontally Oriented Emitters},
   volume = {9},
   url = {https://www.nature.com/articles/s41467-018-05671-x},
   year = {2018}
}

@article{Jeon2018,
   author = {Sohee Jeon and Sunghun Lee and Kyung Hoon Han and Hyun Shin and Kwon Hyeon Kim and Jun Ho Jeong and Jang Joo Kim},
   doi = {10.1002/adom.201701349},
   issn = {21951071},
   journal = {Advanced Optical Materials},
   month = {4},
   publisher = {Wiley-VCH Verlag},
   title = {High-Quality White OLEDs with Comparable Efficiencies to LEDs},
   volume = {6},
   year = {2018}
}

@article{Zhao2024,
   author = {Haonan Zhao and Claire E Arneson and Dejiu Fan and Stephen R Forrest},
   doi = {10.1038/s41586-023-06976-8},
   issn = {1476-4687},
   journal = {Nature},
   month = {2},
   pages = {300-305},
   pmid = {38122821},
   publisher = {Nature Research},
   title = {Stable Blue Phosphorescent Organic Leds That Use Polariton-Enhanced Purcell Effects.},
   volume = {626},
   url = {http://www.ncbi.nlm.nih.gov/pubmed/38122821},
   year = {2024}
}

@article{wang2018rich,
  title={The Rich Photonic World of Plasmonic Nanoparticle Arrays},
  author={Wang, Weijia and Ramezani, Mohammad and V{\"a}kev{\"a}inen, Aaro I and T{\"o}rm{\"a}, P{\"a}ivi and Rivas, Jaime G{\'o}mez and Odom, Teri W},
  journal={Materials today},
  volume={21},
  number={3},
  pages={303--314},
  year={2018},
  publisher={Elsevier}
}

@article{guo2017geometry,
  title={Geometry Dependence of Surface Lattice Resonances in Plasmonic Nanoparticle Arrays},
  author={Guo, Rui and Hakala, Tommi K and T{\"o}rm{\"a}, P{\"a}ivi},
  journal={Physical Review B},
  volume={95},
  number={15},
  pages={155423},
  year={2017},
  publisher={APS}
}

@article{Auguie2008,
   author = {Baptiste Auguié and William L. Barnes},
   doi = {10.1103/PhysRevLett.101.143902},
   issn = {0031-9007},
   journal = {Physical Review Letters},
   month = {9},
   pages = {143902},
   title = {Collective Resonances in Gold Nanoparticle Arrays},
   volume = {101},
   url = {https://link.aps.org/doi/10.1103/PhysRevLett.101.143902},
   year = {2008}
}

@article{Zou2005,
   author = {Shengli Zou and George C. Schatz},
   doi = {10.1016/j.cplett.2004.12.107},
   issn = {00092614},
   journal = {Chemical Physics Letters},
   month = {2},
   pages = {62-67},
   title = {Silver Nanoparticle Array Structures That Produce Giant Enhancements in Electromagnetic Fields},
   volume = {403},
   url = {https://linkinghub.elsevier.com/retrieve/pii/S0009261404020767},
   year = {2005}
}

@article{Kravets2018,
   author = {V. G. Kravets and A. V. Kabashin and W. L. Barnes and A. N. Grigorenko},
   doi = {10.1021/acs.chemrev.8b00243},
   issn = {0009-2665},
   journal = {Chemical Reviews},
   month = {6},
   pages = {5912-5951},
   publisher = {American Chemical Society},
   title = {Plasmonic Surface Lattice Resonances: a Review of Properties and Applications},
   volume = {118},
   url = {https://pubs.acs.org/doi/10.1021/acs.chemrev.8b00243},
   year = {2018}
}

@article{Vecchi2009,
   author = {G. Vecchi and V. Giannini and J. Gómez Rivas},
   doi = {10.1103/PhysRevLett.102.146807},
   issn = {0031-9007},
   journal = {Physical Review Letters},
   month = {4},
   pages = {146807},
   title = {Shaping the Fluorescent Emission by Lattice Resonances in Plasmonic Crystals of Nanoantennas},
   volume = {102},
   url = {https://link.aps.org/doi/10.1103/PhysRevLett.102.146807},
   year = {2009}
}

@article{west2010searching,
  title={Searching for Better Plasmonic Materials},
  author={West, Paul R and Ishii, Satoshi and Naik, Gururaj V and Emani, Naresh K and Shalaev, Vladimir M and Boltasseva, Alexandra},
  journal={Laser \& Photonics Reviews},
  volume={4},
  number={6},
  pages={795--808},
  year={2010},
  publisher={Wiley Online Library}
}

@article{li2021interface,
  title={Interface Engineering in Organic Electronics: Energy-Level Alignment and Charge Transport},
  author={Li, Peicheng and Lu, Zheng-Hong},
  journal={Small Science},
  volume={1},
  number={1},
  pages={2000015},
  year={2021},
  publisher={Wiley Online Library}
}

@article{kanatsiopoulos2025correlation,
  title={Correlation Between Energy-Level Alignment and Interfacial Properties with the OLED Performance},
  author={Kanatsiopoulos, Dimitrios and Papadopoulos, Kyparisis and Tselekidou, Despoina and Kassavetis, Spyros and Logothetidis, Stergious and Gioti, Maria},
  journal={Journal of Physics: Materials},
  year={2025}
}

@article{ding2023recent,
  title={Recent Advances in Linearly Polarized Emission from Organic Light-Emitting Diodes},
  author={Ding, Ran and Ye, Gao-Da and Feng, Jing},
  journal={Applied Physics Letters},
  volume={123},
  number={1},
  year={2023},
  publisher={AIP Publishing}
}

@article{Marcato2025,
   author = {Tommaso Marcato and Jiwoo Oh and Zhan-Hong Lin and Tian Tian and Abhijit Gogoi and Sunil B. Shivarudraiah and Sudhir Kumar and Ananth Govind Rajan and Shuangshuang Zeng and Chih-Jen Shih},
   doi = {10.1038/s41566-025-01785-z},
   issn = {1749-4893},
   journal = {Nature Photonics},
   month = {10},
   title = {Scalable Nanopatterning of Organic Light-Emitting Diodes Beyond the Diffraction Limit},
   url = {https://doi.org/10.1038/s41566-025-01785-z},
   year = {2025}
}

@article{Gubbin2014,
    author = {Gubbin, Christopher R. and Maier, Stefan A. and Kéna-Cohen, Stéphane},
    title = {Low-voltage polariton electroluminescence from an ultrastrongly coupled organic light-emitting diode},
    journal = {Applied Physics Letters},
    volume = {104},
    number = {23},
    pages = {233302},
    year = {2014},
    month = {06},
    issn = {0003-6951},
    doi = {10.1063/1.4871271},
    url = {https://doi.org/10.1063/1.4871271}
}

@article{Abdelmagid2024,
    url = {https://doi.org/10.1515/nanoph-2023-0587},
    title = {Identifying the origin of delayed electroluminescence in a polariton organic light-emitting diode},
    author = {Ahmed Gaber Abdelmagid and Hassan A. Qureshi and Michael A. Papachatzakis and Olli Siltanen and Manish Kumar and Ajith Ashokan and Seyhan Salman and Kimmo Luoma and Konstantinos S. Daskalakis},
    pages = {2565--2573},
    volume = {13},
    number = {14},
    journal = {Nanophotonics},
    doi = {doi:10.1515/nanoph-2023-0587},
    year = {2024}
}

@article{Mischok2023,
    author={Mischok, Andreas and Hillebrandt, Sabina and Kwon, Seonil and Gather, Malte C.},
    title={Highly efficient polaritonic light-emitting diodes with angle-independent narrowband emission},
    journal={Nature Photonics},
    year={2023},
    month={May},
    day={01},
    volume={17},
    number={5},
    pages={393-400},
    issn={1749-4893},
    doi={10.1038/s41566-023-01164-6},
    url={https://doi.org/10.1038/s41566-023-01164-6}
}

@article{Zhang2015,
   author = {Shuyu Zhang and Emiliano R. Martins and Adel G. Diyaf and John I.B. Wilson and Graham A. Turnbull and Ifor D.W. Samuel},
   doi = {10.1016/j.synthmet.2015.03.035},
   issn = {03796779},
   journal = {Synthetic Metals},
   month = {7},
   pages = {127-133},
   publisher = {Elsevier Ltd},
   title = {Calculation of the Emission Power Distribution of Microstructured OLEDs Using the Reciprocity Theorem},
   volume = {205},
   url = {https://linkinghub.elsevier.com/retrieve/pii/S0379677915001526},
   year = {2015}
}

@book{Landau1984,
   author = {L. D. Landau and E. M. Lifshitz},
   edition = {Second edition},
   editor = {E. M. Lifshitz and L. P. Pitaevskii},
   isbn = {0-08-030276-9},
   publisher = {Pergamon Press Ltd.},
   title = {Electrodynamics of Continuous Media},
   year = {1984}
}

@article{Liang2014,
  title={Wideband Analysis of Periodic Structures at Oblique Incidence by Material Independent FDTD Algorithm},
  author={Bingyuan Liang and Ming Bai and Hui Ma and Naiming Ou and Jungang Miao},
  journal={IEEE Transactions on Antennas and Propagation},
  year={2014},
  volume={62},
  pages={354-360},
  url={https://api.semanticscholar.org/CorpusID:31946659}
}

@book{Palik,
      author        = "Palik, Edward D",
      title         = "{Handbook of Optical Constants of Solids I – III}",
      publisher     = "Academic Press",
      address       = "Boston, MA",
      year          = "1997"
}

@article{Li2021,
   author = {Yungui Li and Naresh B. Kotadiya and Bas van der Zee and Paul W. M. Blom and Gert‐Jan A. H. Wetzelaer},
   doi = {10.1002/adom.202001812},
   issn = {2195-1071},
   journal = {Advanced Optical Materials},
   month = {6},
   publisher = {John Wiley and Sons Inc},
   title = {Optical Outcoupling Efficiency of Organic Light‐Emitting Diodes with a Broad Recombination Profile},
   volume = {9},
   url = {https://onlinelibrary.wiley.com/doi/10.1002/adom.202001812},
   year = {2021}
}

@article{Sittinger2022,
    author = {Volker Sittinger and Patricia S. C. Schulze and Christoph Messmer and Andreas Pflug and Jan Christoph Goldschmidt},
    journal = {Opt. Express},
    number = {21},
    pages = {37957--37970},
    publisher = {Optica Publishing Group},
    title = {Complex Refractive Indices of Spiro-TTB and C60 for Optical Analysis of Perovskite Silicon Tandem Solar Cells},
    volume = {30},
    month = {10},
    year = {2022},
    url = {https://opg.optica.org/oe/abstract.cfm?URI=oe-30-21-37957},
    doi = {10.1364/OE.458953},
}

@article{Vos2015,
    author = {Vos, Martijn F. J. and Macco, Bart and Thissen, Nick F. W. and Bol, Ageeth A. and Kessels, W. M. M. (Erwin)},
    title = {Atomic Layer Deposition of Molybdenum Oxide from (NtBu)2(NMe2)2Mo and O2 Plasma},
    journal = {Journal of Vacuum Science \& Technology A},
    volume = {34},
    number = {1},
    pages = {01A103},
    year = {2015},
    month = {09},
    issn = {0734-2101},
    doi = {10.1116/1.4930161},
    url = {https://doi.org/10.1116/1.4930161}
}

@article{Chen2015,
    author = {Chen, Chang-Wen and Hsiao, Sheng-Yi and Chen, Chien-Yu and Kang, Hao-Wei and Huang, Zheng-Yu and Lin, Hao-Wu},
    title = {Optical Properties of Organometal Halide Perovskite Thin Films and General Device Structure Design Rules for Perovskite Single and Tandem Solar Cells},
    journal = {J. Mater. Chem. A},
    year = {2015},
    volume = {3},
    pages = {9152-9159},
    publisher = {The Royal Society of Chemistry},
    doi = {10.1039/C4TA05237D},
    url = {http://dx.doi.org/10.1039/C4TA05237D},
}

@Article{Konig2014,
    author={K{\"o}nig, Tobias A. F.
    and Ledin, Petr A.
    and Kerszulis, Justin
    and Mahmoud, Mahmoud. A.
    and El-Sayed, Mostafa A.
    and Reynolds, John R.
    and Tsukruk, Vladimir V.},
    title={Electrically Tunable Plasmonic Behavior of Nanocube--Polymer Nanomaterials Induced by a Redox-Active Electrochromic Polymer},
    journal={ACS Nano},
    year={2014},
    month={6},
    day={24},
    publisher={American Chemical Society},
    volume={8},
    number={6},
    pages={6182-6192},
    issn={1936-0851},
    doi={10.1021/nn501601e},
    url={https://doi.org/10.1021/nn501601e}
}

@article{Tankeleviciute2024,
   author  = {Egl{\={e}} Tankelevi{\v{c}}i{\={u}}t{\.e} and Ifor D. W. Samuel and Eli Zysman-Colman},
   title   = {The Blue Problem: {OLED} Stability and Degradation Mechanisms},
   journal = {The Journal of Physical Chemistry Letters},
   volume  = {15},
   number  = {4},
   pages   = {1034--1047},
   year    = {2024},
   month   = {2},
   doi     = {10.1021/acs.jpclett.3c03317},
   url     = {https://doi.org/10.1021/acs.jpclett.3c03317},
   publisher = {American Chemical Society}
}

@article{Lee2019,
   author    = {Jiun-Haw Lee and Chia-Hsun Chen and Pei-Hsi Lee and Hung-Yi Lin and Man-kit Leung and Tien-Lung Chiu and Chi-Feng Lin},
   title     = {Blue Organic Light-Emitting Diodes: Current Status, Challenges, and Future Outlook},
   journal   = {Journal of Materials Chemistry C},
   volume    = {7},
   number    = {20},
   pages     = {5874--5888},
   year      = {2019},
   doi       = {10.1039/C9TC00204A},
   url       = {http://dx.doi.org/10.1039/C9TC00204A},
   publisher = {The Royal Society of Chemistry},
   issn      = {2050-7526}
}

@article{Tang1987,
   author = {C. W. Tang and S. A. VanSlyke},
   doi = {10.1063/1.98799},
   issn = {0003-6951},
   journal = {Applied Physics Letters},
   month = {9},
   pages = {913-915},
   title = {Organic Electroluminescent Diodes},
   volume = {51},
   url = {https://pubs.aip.org/apl/article/51/12/913/1023099/Organic-electroluminescent-diodes},
   year = {1987}
}

@article{Chan2021,
   author = {Chin-Yiu Chan and Masaki Tanaka and Yi-Ting Lee and Yiu-Wing Wong and Hajime Nakanotani and Takuji Hatakeyama and Chihaya Adachi},
   doi = {10.1038/s41566-020-00745-z},
   issn = {1749-4885},
   journal = {Nature Photonics},
   month = {3},
   pages = {203-207},
   publisher = {Nature Research},
   title = {Stable Pure-Blue Hyperfluorescence Organic Light-Emitting Diodes with High-Efficiency and Narrow Emission},
   volume = {15},
   url = {https://www.nature.com/articles/s41566-020-00745-z},
   year = {2021}
}

@article{Hatakeyama2016,
   author = {Takuji Hatakeyama and Kazushi Shiren and Kiichi Nakajima and Shintaro Nomura and Soichiro Nakatsuka and Keisuke Kinoshita and Jingping Ni and Yohei Ono and Toshiaki Ikuta},
   doi = {10.1002/adma.201505491},
   issn = {0935-9648},
   journal = {Advanced Materials},
   month = {4},
   pages = {2777-2781},
   publisher = {Wiley-VCH Verlag},
   title = {Ultrapure Blue Thermally Activated Delayed Fluorescence Molecules: Efficient Homo–Lumo Separation by the Multiple Resonance Effect},
   volume = {28},
   url = {https://onlinelibrary.wiley.com/doi/10.1002/adma.201505491},
   year = {2016}
}

@article{Adachi2001,
   author = {Chihaya Adachi and Marc A. Baldo and Mark E. Thompson and Stephen R. Forrest},
   doi = {10.1063/1.1409582},
   issn = {0021-8979},
   journal = {Journal of Applied Physics},
   month = {11},
   pages = {5048-5051},
   title = {Nearly 100\% Internal Phosphorescence Efficiency in an Organic Light-Emitting Device},
   volume = {90},
   url = {https://pubs.aip.org/jap/article/90/10/5048/181142/Nearly-100-internal-phosphorescence-efficiency-in},
   year = {2001}
}

@article{Wu2018,
   author = {Tien-Lin Wu and Min-Jie Huang and Chih-Chun Lin and Pei-Yun Huang and Tsu-Yu Chou and Ren-Wu Chen-Cheng and Hao-Wu Lin and Rai-Shung Liu and Chien-Hong Cheng},
   doi = {10.1038/s41566-018-0112-9},
   issn = {1749-4885},
   journal = {Nature Photonics},
   month = {4},
   pages = {235-240},
   publisher = {Nature Publishing Group},
   title = {Diboron Compound-Based Organic Light-Emitting Diodes with High Efficiency and Reduced Efficiency Roll-Off},
   volume = {12},
   url = {https://www.nature.com/articles/s41566-018-0112-9},
   year = {2018}
}

@misc{Comsol,
    title = {{COMSOL Multiphysics® v. 6.3}},
    howpublished = {\url{www.comsol.com}},
    author = {COMSOL AB, Stockholm, Sweden}
}

@misc{ComsolModeTracking,
    author = {Yuanshen Li},
    howpublished = {\url{https://www.comsol.com/blogs/tracking-eigenmodes-over-parameteric-sweeps}},
    note = {Accessed: 2025-07-17},
    year = {2024},
    title = {{COMSOL} Blog: {T}racking Eigenmodes Over Parametric Sweeps}
}

@misc{ComsolModeAnalysis,
    author = {Sergey Yankin},
    howpublished = {\url{https://www.comsol.com/blogs/mode-analysis-for-electromagnetic-waveguides-in-comsol}},
    note = {Accessed: 2025-07-17},
    year = {2022},
    title = {{COMSOL} Blog: {M}ode Analysis for Electromagnetic Waveguides In {COMSOL}®}
}

@inproceedings{10.1117/12.2649342,
author = {Jaime G{\'o}mez Rivas and Mohammad Ramezani and Marc Verschuuren and Gabriel Castellanos},
title = {Large-area Scatterometry for Nanoscale Metrology},
volume = {12433},
booktitle = {Advanced Fabrication Technologies for Micro/Nano Optics and Photonics Xvi},
editor = {Georg von Freymann and Eva Blasco and Debashis Chanda},
organization = {International Society for Optics and Photonics},
publisher = {SPIE},
pages = {124330I},
year = {2023},
doi = {10.1117/12.2649342},
URL = {https://doi.org/10.1117/12.2649342}
}

@article{Nevels_2022,
doi = {10.1088/2040-8986/ac7abe},
url = {https://doi.org/10.1088/2040-8986/ac7abe},
year = {2022},
month = {7},
publisher = {IOP Publishing},
volume = {24},
number = {9},
pages = {094002},
author = {Nevels, Teun D G and Ruijs, Lieke J M and van de Meugheuvel, Paul and Verschuuren, Marc A and Rivas, Jaime Gómez and Ramezani, Mohammad},
title = {Novel Optical Metrology for Inspection of Nanostructures Fabricated by Substrate Conformal Imprint Lithography},
journal = {Journal of Optics}
}

@article{doi:10.1021/nn501601e,
author = {K{\"o}nig, Tobias A. F. and Ledin, Petr A. and Kerszulis, Justin and Mahmoud, Mahmoud. A. and El-Sayed, Mostafa A. and Reynolds, John R. and Tsukruk, Vladimir V.},
title = {Electrically Tunable Plasmonic Behavior of Nanocube–Polymer Nanomaterials Induced by a Redox-Active Electrochromic Polymer},
journal = {ACS Nano},
volume = {8},
number = {6},
pages = {6182-6192},
year = {2014},
doi = {10.1021/nn501601e},
}

@article{Rakic:95,
author = {Aleksandar D. Raki\'{c}},
journal = {Appl. Opt.},
number = {22},
pages = {4755--4767},
publisher = {Optica Publishing Group},
title = {Algorithm for the Determination of Intrinsic Optical Constants of Metal Films: Application to Aluminum},
volume = {34},
month = {8},
year = {1995},
url = {https://opg.optica.org/ao/abstract.cfm?URI=ao-34-22-4755},
doi = {10.1364/AO.34.004755},
}

@book{Adams1981,
   author = {M. J. Adams},
   isbn = {0471279692},
   publisher = {John Wiley \& Sons Ltd.},
   title = {An Introduction to Optical Waveguides},
   year = {1981},
   pages = {61--68}
}

@article{Murai2013,
   author = {S. Murai and M. A. Verschuuren and G. Lozano and G. Pirruccio and S. R. K. Rodriguez and J. Gómez Rivas},
   doi = {10.1364/OE.21.004250},
   issn = {1094-4087},
   journal = {Optics Express},
   month = {2},
   pages = {4250},
   title = {Hybrid Plasmonic-Photonic Modes in Diffractive Arrays of Nanoparticles Coupled to Light-Emitting Optical Waveguides},
   volume = {21},
   url = {https://opg.optica.org/oe/abstract.cfm?uri=oe-21-4-4250},
   year = {2013}
}

@misc{Lehikoinen2026,
  author = {Joel Lehikoinen and Rebecca Heilmann and Aron J. J. Dahlberg and Eero Härmä and Malek Mahmoudi and Arpan Dutta and Konstantinos S. Daskalakis and Päivi Törmä},
  title  = {Flat Bands from Diffraction in Periodic Systems},
  note   = {Submitted. arXiv: 2602.21830},
  year = {2026},
  url = {https://arxiv.org/abs/2602.21830}
}

@misc{Heilmann2026,
  author = {Rebecca Heilmann and Joel Lehikoinen and Sioneh Eyvazi and Evgeny A. Mamonov and Päivi Törmä},
  title  = {Chains of nanoparticles for flat-band emission and lasing},
  note   = {Submitted. arXiv: 2602.21825},
  year = {2026},
  url = {https://arxiv.org/abs/2602.21825},
  eprint = {2602.21825},
  archivePrefix= {arXiv},
  primaryClass = {physics.optics}
}

@article{sun2008enhanced,
  title={Enhanced Light Out-Coupling of Organic Light-Emitting Devices Using Embedded Low-Index Grids},
  author={Sun, Yiru and Forrest, Stephen R},
  journal={Nature Photonics},
  volume={2},
  number={8},
  pages={483--487},
  year={2008},
  publisher={Nature Publishing Group UK London}
}

@article{chang2013nano,
  title={Nano-particle Based Scattering Layers for Optical Efficiency Enhancement of Organic Light-Emitting Diodes and Organic Solar Cells},
  author={Chang, Hong-Wei and Lee, Jonghee and Hofmann, Simone and Hyun Kim, Yong and M{\"u}ller-Meskamp, Lars and L{\"u}ssem, Bj{\"o}rn and Wu, Chung-Chih and Leo, Karl and Gather, Malte C},
  journal={Journal of Applied Physics},
  volume={113},
  number={20},
  year={2013},
  publisher={AIP Publishing}
}

@article{ou2014extremely,
  title={Extremely Efficient White Organic Light-Emitting Diodes for General Lighting},
  author={Ou, Qing-Dong and Zhou, Lei and Li, Yan-Qing and Shen, Su and Chen, Jing-De and Li, Chi and Wang, Qian-Kun and Lee, Shuit-Tong and Tang, Jian-Xin},
  journal={Advanced Functional Materials},
  volume={24},
  number={46},
  pages={7249--7256},
  year={2014},
  publisher={Wiley Online Library}
}

@article{zhou2015efficiently,
  title={Efficiently Releasing the Trapped Energy Flow in White Organic Light-Emitting Diodes with Multifunctional Nanofunnel Arrays},
  author={Zhou, Lei and Ou, Qing-Dong and Li, Yan-Qing and Xiang, Heng-Yang and Xu, Lu-Hai and Chen, Jing-De and Li, Chi and Shen, Su and Lee, Shuit-Tong and Tang, Jian-Xin},
  journal={Advanced Functional Materials},
  volume={25},
  number={18},
  pages={2660--2668},
  year={2015},
  publisher={Wiley Online Library}
}

@article{lee2016effect,
  title={The Effect of Localized Surface Plasmon Resonance on the Emission Color Change in Organic Light Emitting Diodes},
  author={Lee, Illhwan and Park, Jae Yong and Hong, Kihyon and Son, Jun Ho and Kim, Sungjun and Lee, Jong-Lam},
  journal={Nanoscale},
  volume={8},
  number={12},
  pages={6463--6467},
  year={2016},
  publisher={Royal Society of Chemistry}
}

@article{chen2024dual,
  title={Dual-Color Emissive OLED with Orthogonal Polarization Modes},
  author={Chen, Ruixiang and Liang, Ningning and Zhai, Tianrui},
  journal={Nature Communications},
  volume={15},
  number={1},
  pages={1331},
  year={2024},
  publisher={Nature Publishing Group UK London}
}

@article{zhou2013lasing,
  title={Lasing action in strongly coupled plasmonic nanocavity arrays},
  author={Zhou, Wei and Dridi, Montacer and Suh, Jae Yong and Kim, Chul Hoon and Co, Dick T and Wasielewski, Michael R and Schatz, George C and Odom, Teri W},
  journal={Nature nanotechnology},
  volume={8},
  number={7},
  pages={506--511},
  year={2013},
  publisher={Nature Publishing Group UK London}
}

@article{freire2025plasmonic,
  title={Plasmonic lattice lasers},
  author={Freire-Fern{\'a}ndez, Francisco and Park, Sang-Min and Tan, Max JH and Odom, Teri W},
  journal={Nature Reviews Materials},
  volume={10},
  number={8},
  pages={604--616},
  year={2025},
  publisher={Nature Publishing Group UK London}
}

@article{hakala2018bose,
  title={Bose--Einstein condensation in a plasmonic lattice},
  author={Hakala, Tommi K and Moilanen, Antti J and V{\"a}kev{\"a}inen, Aaro I and Guo, Rui and Martikainen, Jani-Petri and Daskalakis, Konstantinos S and Rekola, Heikki T and Julku, Aleksi and T{\"o}rm{\"a}, P{\"a}ivi},
  journal={Nature Physics},
  volume={14},
  number={7},
  pages={739--744},
  year={2018},
  publisher={Nature Publishing Group UK London}
}

@misc{Dahlberg2026Enhancement,
  author = {Dahlberg, Aron J. J. and Gužauskas, Matas and Mahmoudi, Malek and Lehikoinen, Joel and Volyniuk, Dmytro and Kumar, Manish and Berghuis, Anton Matthijs and Qureshi, Hassan A. and Keruckiene, Rasa and Papachatzakis, Michael A. and Mamonov, Evgeny A. and Heilmann, Rebecca and Lyyra, Henri and Siltanen, Olli and Barnes, William L. and Gómez Rivas, Jaime and Daskalakis, Konstantinos S. and Gražulevičius, Juozas Vidas and T{\"o}rm{\"a}, P{\"a}ivi },
  title={Dataset for the paper: "Nanoparticle Arrays for Efficient Organic Light-Emitting Diode Emission Management"},
  year   = {2026},
  note   = {Zenodo, DOI: https://doi.org/10.5281/zenodo.19063111}
}

\clearpage
\section*{Supporting Information}
\addcontentsline{toc}{section}{Supporting Information}

\renewcommand{\thefigure}{S\arabic{figure}}
\renewcommand{\thetable}{S\arabic{table}}
\renewcommand{\thesection}{S\arabic{section}}
\setcounter{figure}{0}

\subsection*{Light extraction from inactive pixel}

\begin{figure}[H]
\includegraphics[width=\textwidth]{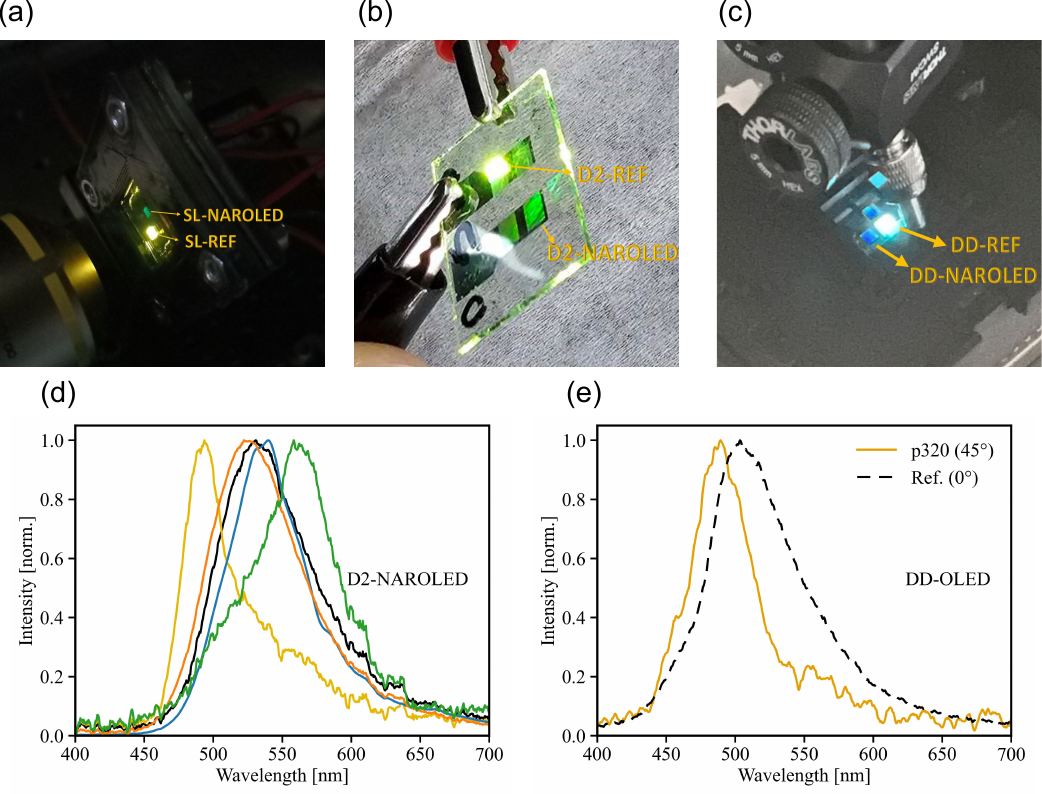}
\caption{Photographs of (a) SL-OLED, (b) D2-OLED, and (c) DD-OLED, showing inactive pixels containing nanostructures that emit light when a neighboring pixel is active. (d) EL spectra of the inactive D2-NAROLED recorded at different viewing angles; and (e) the emission spectra of the inactive DD-NAROLED and the active DD-REF.}
\label{fig:SI-InactivePixels}
\end{figure}

Figure~\ref{fig:SI-InactivePixels} demonstrates light extraction from inactive nanoarrayed pixels in SL-OLED, D2-OLED, and DD-OLED devices. As shown in panels (a–c), the inactive NAROLEDs pixels emit light when a neighboring reference pixel is electrically driven, despite not being biased themselves. Panels (d) and (e) present the corresponding EL spectra: (d) shows the angular-dependent emission of the inactive D2-NAROLED, while (e) compares the spectra of the inactive DD-NAROLED and the active DD-REF. This observation provides direct evidence of passive optical coupling, where emission from the active pixel is coupled into waveguided modes and subsequently outcoupled by the nanoarrays.

\newpage
\subsection*{Metrology of the nanoparticle arrays on ITO}
\label{SI_metrology}
\begin{figure}[H]
\includegraphics[width=0.95\textwidth]{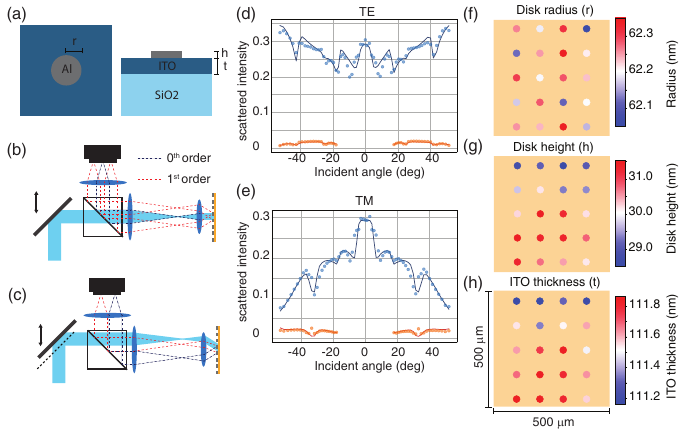}
\caption{(a) Schematic top view and cross section of a unit cell of the nanoparticle array as used in the RCWA model. (b) Schematic of the scatterometer for normal incidence illumination, and (c) for illumination under an off-normal angle. (d)-(e) Angle resolved measurements of the 0\textsuperscript{th}-order (blue dots) and $\pm$1\textsuperscript{st}-orders (red dots) of diffraction for TE (d) and TM (e) polarization. The solid blue curves and red curves are the RCWA fits to the data. (f)-(g) Maps of the retrieved dimensions at 20 positions on the 500~\textmu m x 500~\textmu m array: disk radius (f), disk height (g), and ITO thickness (h).}
\label{fig:SI-Sample_Metrology}
\end{figure}

To verify the quality and homogeneity of the nanoparticle arrays, they have been examined in a commercial scatterometry tool (LabScatter, TeraNova). To illustrate the homogeneity of the arrays, we have measured the array with a period of 370 nm before depositing the active layers, as schematically shown in Figure~\ref{fig:SI-Sample_Metrology} (a). The scatterometer works by focusing a laser beam (405 nm) into the back focal plane (BFP) of an objective lens, resulting in a collimated beam at the front focal plane with a diameter of $\sim$10 ~\textmu m. The specularly reflected light (0th order) and the diffracted light (1st order) are collected through the same objective lens and detected by a camera (Figure~\ref{fig:SI-Sample_Metrology} (b)). By scanning the position of the focused beam in the BFP, the angle of incidence of the laser beam onto the sample is varied from -55 to +55 degrees with respect to normal incidence (Figure~\ref{fig:SI-Sample_Metrology} (c)).  

The scattered intensity, normalized to the reflection of a polished undoped silicon wafer, is plotted for TE and TM polarized light in Figures ~\ref{fig:SI-Sample_Metrology} (d) and (e), respectively. The blue circles are the detected intensity of the 0\textsuperscript{th} order (specular reflection) and the orange circles the 1\textsuperscript{st} and -1\textsuperscript{st} diffraction orders. The dimensions of the nanoparticle array in the illuminated area are retrieved by fitting the angle-dependent intensity of the different orders with calculations using the rigorous coupled wave analysis (RCWA) method~\cite{Nevels_2022,10.1117/12.2649342}. The refractive index of the ITO is obtained from K\"onig et. al.\cite{doi:10.1021/nn501601e} and the refractive index of the aluminum from R\'akic \cite{Rakic:95}. The particle radius ($r$), particle height ($h$), and ITO thickness ($t$) are free parameters that are fitted to the data. The resulting fits at one position in the array for TE and TM polarizations are shown with the blue and orange curves in Figs.~\ref{fig:SI-Sample_Metrology} (d) and (e). While the fits for both polarizations show the same features as the measurements, there are small quantitative discrepancies. These discrepancies can be attributed to the fact that the RCWA model calculates the response for a perfect array without surface roughness or rounding of the edges of the particle. 

Next, we have performed an area scan over the 500~\textmu m x 500~\textmu m nanoparticle array by measuring in a matrix of 4x5 points. For each measurement, we extract the $r$, $h$, and $t$. The results are plotted in Figs.~\ref{fig:SI-Sample_Metrology} (f-h). The retrieved dimensions show excellent homogeneity over the sample area, with the following dimensions and spatial variation over the different positions: $r=62.1\pm 0.1$ nm$, h=30.2\pm 0.9$ nm, and $t=111\pm0.3$ nm. The measured spatial variations of the dimensions are on the order of the sensitivity of the scatterometer. It should be noted that using different permittivities for ITO and Al could give slightly different absolute values of the dimensions; however, this would not significantly affect the spatial variation of the dimensions.  Despite the excellent homogeneity of the sample, there appears to be a systematic increase in the disk height towards the bottom of the array (Figure ~\ref{fig:SI-Sample_Metrology} (g)), which correlates with the increase in the ITO thickness (Figure ~\ref{fig:SI-Sample_Metrology} (h)).  

\newpage
\subsection{Electrical Stability of NAROLEDs}
The electrical stability of the NAROLEDs was studied by recording the JV-curves and comparing them to reference OLEDs. The JV-curves are presented in logarithmic and linear scale (inset), respectively.

\begin{figure}[H]
\includegraphics[width=\textwidth]{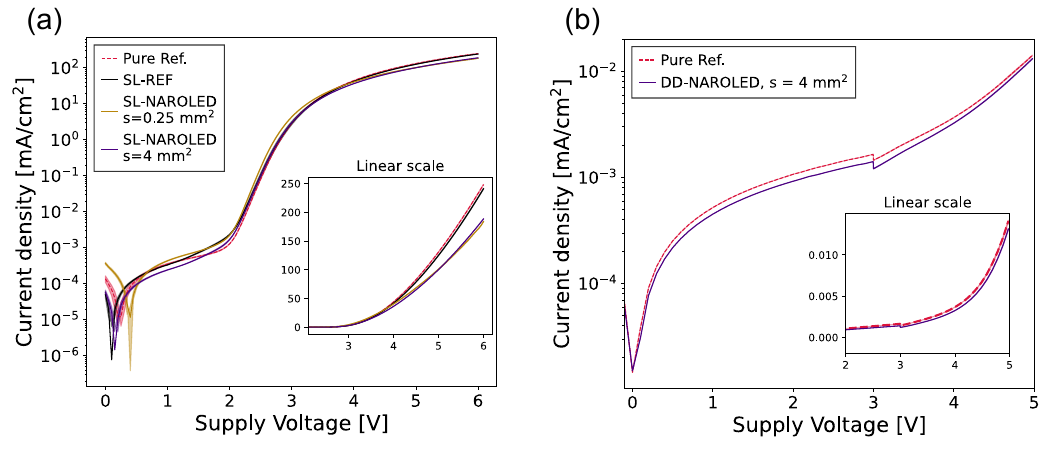}
\caption{Current density in logarithmic scale and linear scale (inset) at different supply voltages for (a) SL-OLEDs and (b) DD-OLEDs. Panel (a) shows a standard SL-OLED without nanopatterns (red dashed), SL-REF pixel (black), SL-NAROLED with a period of 320 nm, diameter of 120 nm with an array size of 0.25 mm$^2$ (golden), and SL-NAROLED with a period of 350 nm, diameter of 120 nm with an array size of 4 mm$^2$ (indigo). Similarly, in (b), the pure reference (red dashed) and a DD-NAROLED (indigo) with a period of 300 nm, a diameter of 100 nm, with an array size of 4 mm$^2$ are shown. Standard errors were obtained from (a) five and (b) three repeated measurements and are shown by the thickness of each curve.}
\label{fig:SI - JV Curves}
\end{figure}

In Figure~\ref{fig:SI - JV Curves}(a), the JV-sweeps of a pure reference (red dashed), i.e a pixel in a substrate that has not gone through nanofabrication, SL-REF (black), and SL-NAROLED with period of 320 nm, diameter of 120 nm with an array size of 0.25 mm$^2$ (golden) and SL-NAROLED with a period of 350 nm, diameter of 120 nm with an array size of 4 mm$^2$ (indigo). The standard error, shown as the thickness of each curve, was calculated from five repeated measurements per pixel. Comparing the pure reference and SL-REF, it can be noted that the nanofabrication processing does not reduce the pixel quality. The current density of the SL-NAROLED follows that of the SL-REF through the injection-limited regime, after which it starts to deviate from the reference at high voltages, irrespective of the area of the nanostructure within the pixel. 

Similarly, Figure~\ref{fig:SI - JV Curves}(b) shows the JV-characteristics of DD-OLEDs. The measurements were repeated over three independent JV cycles to verify reproducibility, evaluate potential degradation effects, and calculate the standard error. The current density of the pure reference pixel (red dashed) is only slightly higher than that of the DD-NAROLED with a period of 300 nm, a diameter of 100 nm, with an array size of 4 mm$^2$ (indigo) in the measured voltage range. This indicates that the DD-NAROLED is stable as well.
The sweeps captured here are within the early injection-limited region, where the SL-NAROLED exhibited similar operation compared to its respective reference. These measurements indicate the viability, from an electrical standpoint, of embedding OLEDs with nanostructures.

\newpage
\subsection{Further Studies of the NAROLEDs}
\begin{figure}[H]
\includegraphics[width=\textwidth]{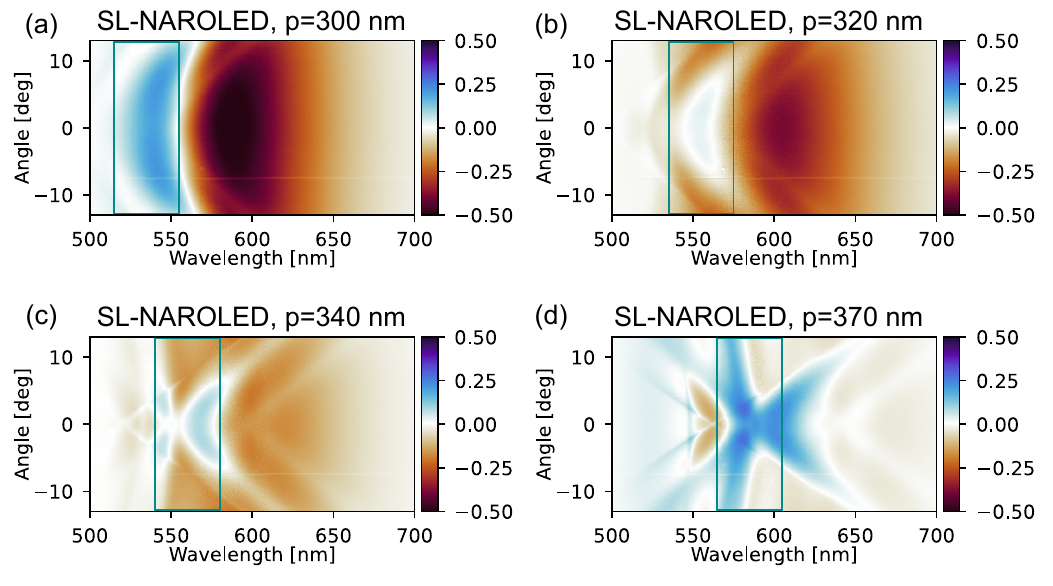}
\caption{Angularly-resolved spectrum for (a) p=300 nm, (b) p=320 nm, (c) p=340 nm, and (d) p=370 nm  at 250 mA/cm$^2$. The intensity was normalized relative to the SL-REF according to Equation~\ref{e:reduced_intensity}. The dark cyan boxes represent 40 nm wide windows within the maximal enhancement region of each SL-NAROLED, respectively.}
\label{fig:SI-Aalto_all_dispersions}
\end{figure}

In Figure~\ref{fig:SI-Aalto_all_dispersions}, the EL spectrum of the SL-NAROLED was studied with square geometries of periods p = [300 nm, 320 nm, 340 nm, 370 nm]. The colorscale represents normalization with respect to the SL-REF according to Equation~\ref{e:reduced_intensity}. Accordingly, the first DO mode location redshifts with larger periods, with p=370 nm corresponding to SLR mode overlap with the emission maximum. Here, the dark-cyan box represents a narrow 40 nm wide window around the maximum enhancement regions of each SL-NAROLED, respectively. We consider it to be representative of narrow emitters, whose spectral overlap could be better tuned with the SLR modes.

\begin{figure}[H]
\includegraphics[width=\textwidth]{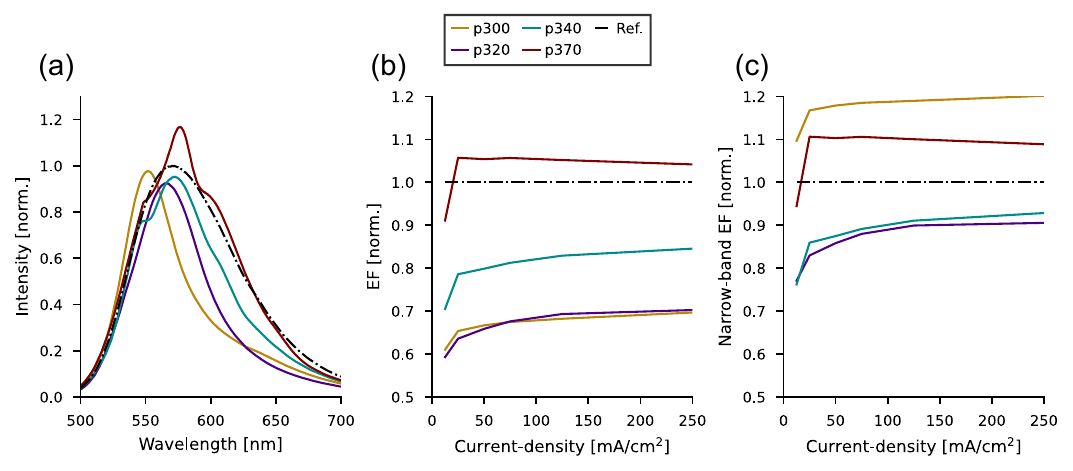}
\caption{Integrated parameters of the angularly-resolved spectra of the SL-NAROLEDs in~\ref{fig:SI-Aalto_all_dispersions} with square array periodicities of p=300 nm (golden), p=320 nm (indigo), p=340 nm (dark cyan), and p=370 nm (maroon), normalized to the respective SL-REF (black dash-dotted) parameter.  (a) Spectra of different SL-NAROLEDs at a constant current density of 250 mA/cm$^2$. EF from integrated intensity counts as a function of current density for (b) the entire measured angle-wavelength space, and (c) EF from a 40 nm wide wavelength band around the maximum enhancement, shown as the dark cyan rectangles in~\ref{fig:SI-Aalto_all_dispersions}(a-d). }
\label{fig:SI-Aalto_Integrated}
\end{figure}

\begin{figure}[H]
\begin{subfigure}{\textwidth}
\includegraphics[width=\textwidth]{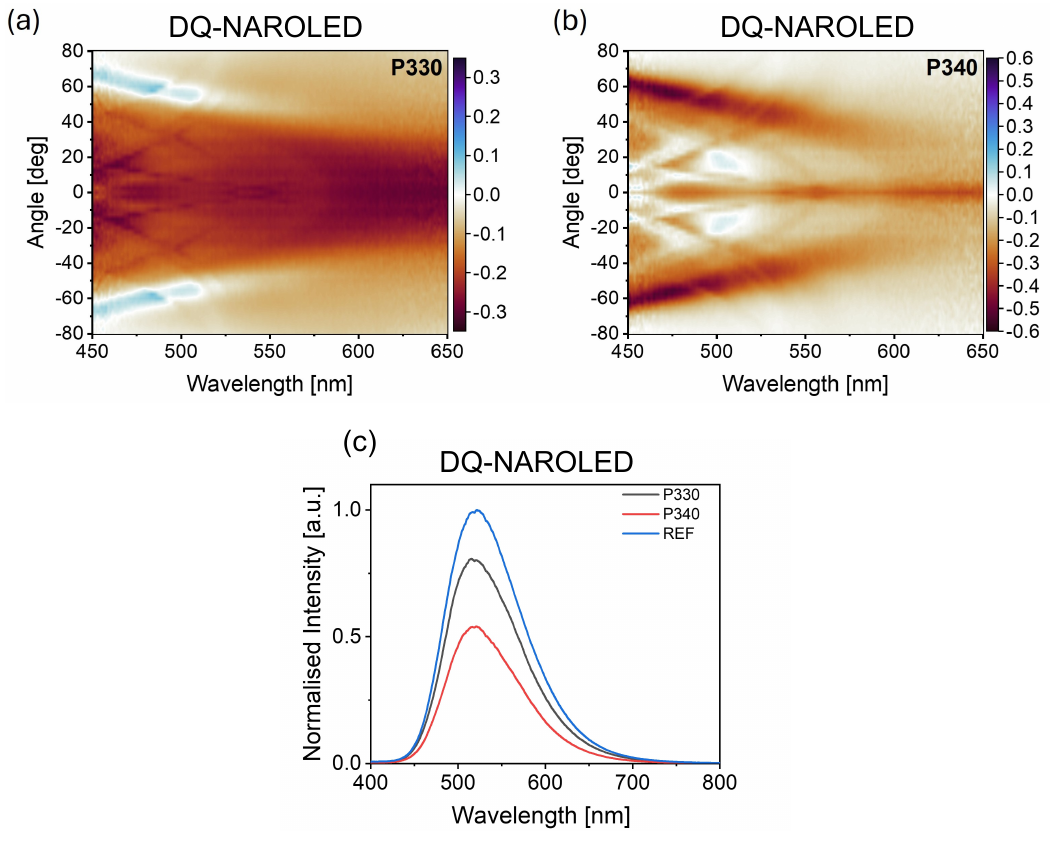}
\end{subfigure}
\caption{Angularly-resolved spectrum for (a) p=330 nm, (b) p=340 nm, at 2.5 mA/cm$^2$. The intensity was normalized relative to the DQ-REF according to Equation~\ref{e:reduced_intensity} of the main text. (c) Integrated electroluminescence spectra from the raw data of DQ-NAROLEDs of (a,b) and DQ-REF.}
\label{fig:SI-KTU_All_periods}
\end{figure}

\begin{figure}[H]
\begin{subfigure}{0.3\textwidth}
\includegraphics[width=\textwidth]{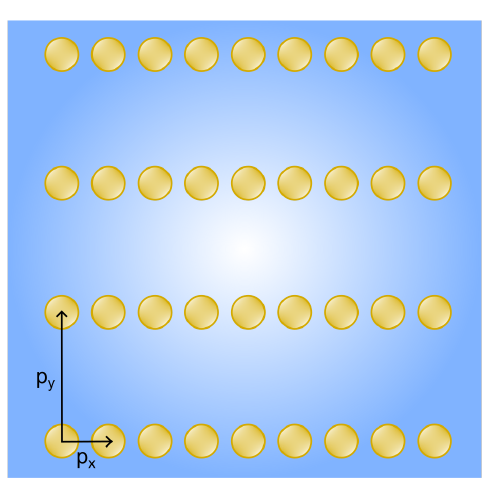}
\end{subfigure}
\caption{Schematic of a chain lattice, where the $p_x$ and $p_y$ form a rectangular geometry with a high aspect ratio, producing flat bands. For the flat band presented in Figure~\ref{fig:3}(b,d) the aspect ratio of $p_x$:$p_y$ was 8.1, i.e $p_y$ = $8.1\times p_x$.}
\label{fig:SI-Chainlattice Schematic}
\end{figure}

\begin{figure}[H]
\begin{subfigure}{\textwidth}
\includegraphics[width=\textwidth]{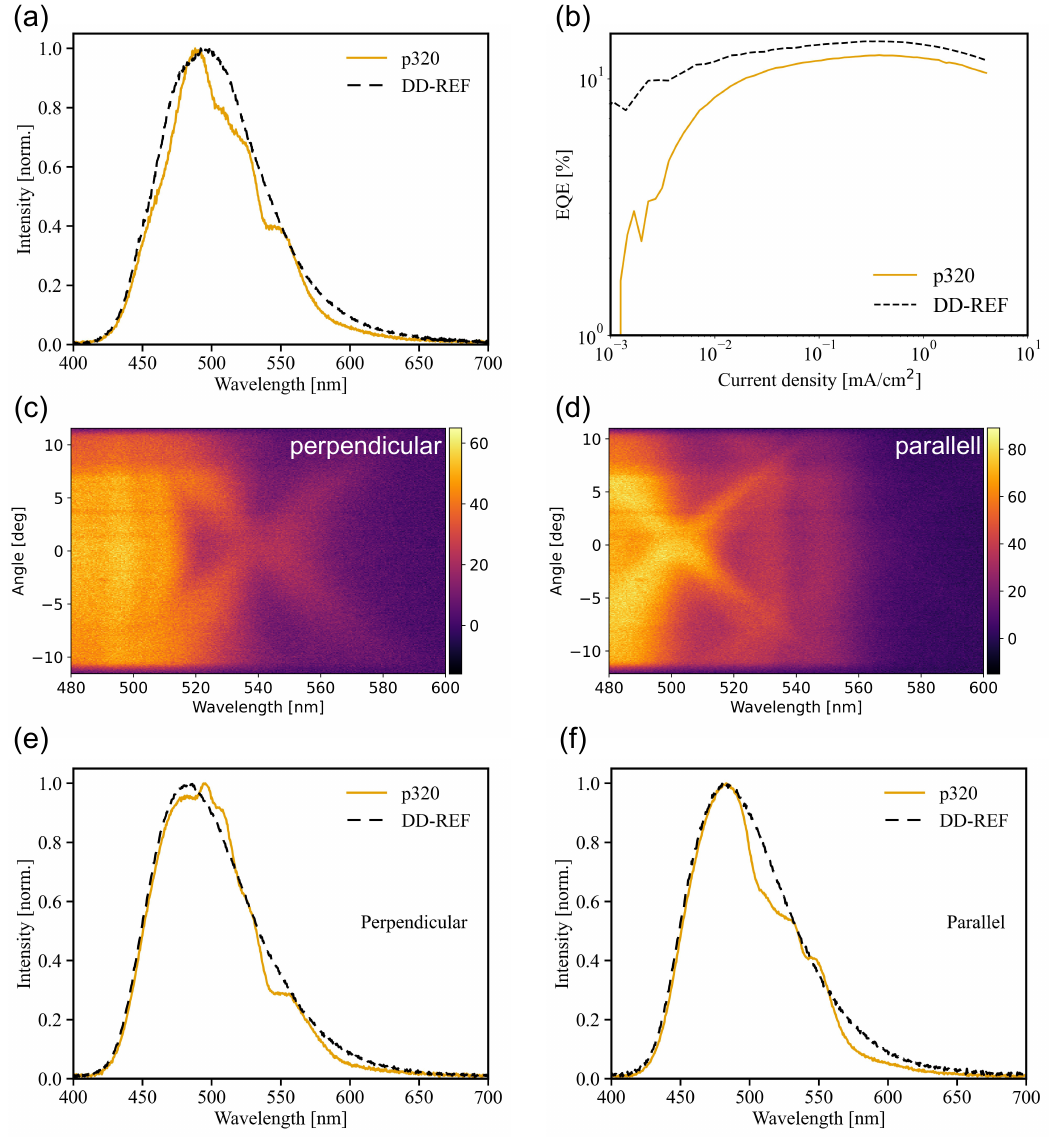}
\end{subfigure}
\caption{(a) EL spectra and (b) EQE vs current density curves of DD-NAROLEDs measured under an applied voltage ranging from 0 to 10 V, comparing a reference pixel with a nanostructured pixel (period: 320 nm, diameter: 100 nm, height: 30 nm, array area: $2~\text{mm} \times 2~\text{mm}$). (c) Perpendicular and (d) parallel-polarized angle-resolved EL spectra of the same OLED measured over emission angles of $\pm \ang{11.54}$, using a Fourier imaging setup (150~\textmu m slit width and 2 s exposure time). (e) Normalized angle-integrated perpendicular and (f) parallel-polarized EL spectra extracted from the Fourier spectroscopy measurements.}
\label{fig:SI-UTU-measurements}
\end{figure}

The spectral shape and the enhancement factor (EF) as a function of current density of the SL-NAROLEDs of Figure~\ref{fig:SI-Aalto_all_dispersions} are shown in Figure~\ref{fig:SI-Aalto_Integrated}. The peak position and shape of the emission spectrum can be tuned with a simple modification of the period of the square array. The total intensity, integrated from the entire measured region of $\pm 13 \degree$ and 500-700 nm, is shown in Figure~\ref{fig:SI-Aalto_Integrated}(b). The array with period p=370 nm is the only structure showing overall enhancement. This is largely due to the competition of enhanced mode strength and losses at non-overlapping modes. However, considering emitters with only a narrow emission band around the respective maximum enhancements, \textit{i.e} the dark cyan boxes in Figure~\ref{fig:SI-Aalto_all_dispersions}, the EF of period p=300 nm increases from $\sim 0.7$ to 1.2 as shown in Figure~\ref{fig:SI-Aalto_Integrated}(c). This gives a clear indication that using narrow emitters is a viable option for achieving significant enhancements.

% KTU FIGURES

\subsection{Identifying array modes in experimental data}

In the main text, we mainly rely on visual comparison to associate the features of experimental observations with the SLRs of the array; see, e.g., the discussion concerning Figs. 2 and 3 of the main text. Here, we provide a more in-depth explanation of how the theoretically calculated mode dispersions are linked to the experimental data. There are two key challenges in associating SLR modes to peaks in experimental data: First, the expected spectral positions of the SLR modes calculated from Eq.~(2) consider only the spectral position of the DO; however, the SLR is a hybridization of the broad LSPR and the narrow DO. Consequently, SLRs appear as Fano resonances, i.e., asymmetric resonances, where the shape of the resonance depends on the spectral position of the DO relative to the LSPR~\cite{Kravets2018}. Thus, we expect to see an extremum near the spectral position of the DO; but the extremum may be a maximum or a minimum and it may be shifted from the spectral position of the DO. Second, in computing the empty-lattice dispersions of Figs.~2 and~3 of the main text, we assumed that the data were recorded along the line $k_x = 0$. However, any tilt of the sample about the $y$ axis (the axis of the spectrometer slit) $\theta_x$ causes the data to be recorded along $k_x = k_0 \sin \theta_x$. Inserting this into the dispersion [Eq.~(2) of the main text] of the first order TM modes $(\pm 1, 0)$ yields
\begin{equation}
    E(k_y) = \frac{\hbar c_0}{n} \sqrt{\left(k_0 \sin \theta_x \pm \frac{2 \pi}{p}\right)^2 + k_y^2},
\end{equation}
where the plus (minus) sign corresponds to the DO with $m_1 = 1$ ($m_1 = -1$). Sample tilt thus lifts the degeneracy of the TM modes $(\pm 1, 0)$ and shifts them in energy (one mode to a higher, the other to a lower energy) – this is best visible in Fig.~3(a) of the main text for the modes TE$_0^{(\pm 1, 0)}$. 
% So based on the theory we don't know (1) where the peak is going to be, (2) if it will be a minimum or a maximum, and (3) how many peaks there are...

\begin{figure}
    \centering
    \includegraphics[width=\linewidth]{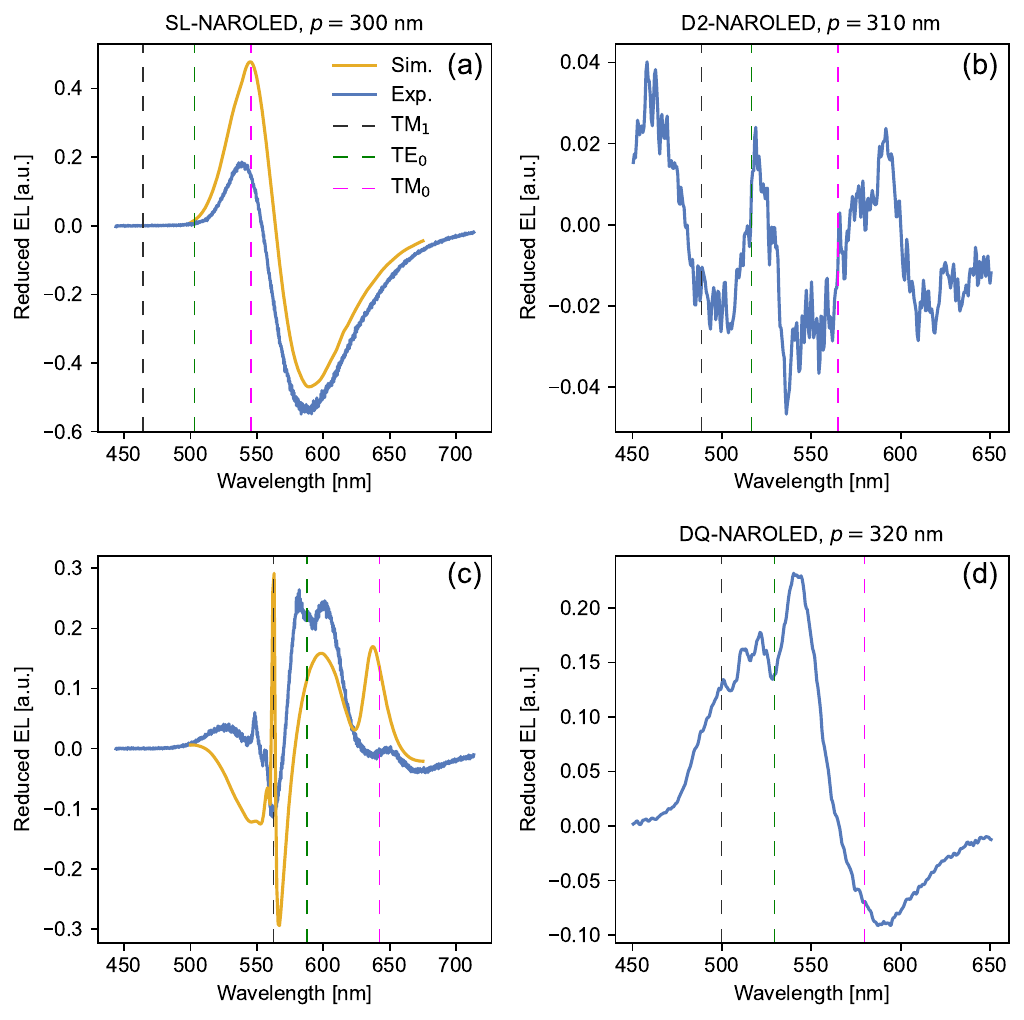}
    \caption{Cuts of the experimental (blue lines) and simulated (gold lines) electroluminescence data [reduced intensity, see Eq.~(1) of main text] presented in Figs.~2(a, b, d, e) of the main text along the line $\theta = 0\degree{}$. The dashed vertical lines indicate the spectral positions of the crossings of the SLRs associated with diffraction order $(\pm 1, 0)$ and $(0, \pm 1)$ calculated using Eq.~(3) of the main text and the effective refractive indices of the waveguide modes (see Fig.~5 of the main text).}
    \label{fig:SI-0deg-cuts}
\end{figure}

Figure~\ref{fig:SI-0deg-cuts} shows cuts of the EL data presented in Fig.~2 of the main text and the spectral positions of the crossings of the SLRs associated with first diffraction orders calculated using Eq.~(3) of the main text and the effective refractive indices of the waveguide modes (see Fig.~5 of main text) for the SL (two array periods), DD, and DQ devices. The spectral positions have been calculated from Eq.~(3) of the main text. Because the effective index of the waveguide mode is a function of wavelength, the equation must be solved numerically; we used fixed-point iteration.

For SL-NAROLED with square arrays with a period of 300~nm [see Fig.~\ref{fig:SI-0deg-cuts}(a)], the spectral positions of the modes TE$_0^{(\pm 1, 0)}$, TE$_0^{(0, \pm 1)}$, TM$_1^{(\pm 1, 0)}$, and TM$_1^{(0, \pm 1)}$ fall outside the emission spectrum of the emitter. The spectral positions of the modes TM$_0^{(\pm 1, 0)}$ and TM$_0^{(0, \pm 1)}$ coincide with an EL maximum. When the array period is changed to 370~nm [Fig.~\ref{fig:SI-0deg-cuts}(c)], all of the calculated SLR wavelengths are close to a minimum or a maximum of the EL. Notably, a double peak appears around the spectral position corresponding to the TE$_0$ waveguide mode (indicated by the green dashed line); we take this to be a signature of the sample tilt discussed above. Similarly, the peak near 550~nm corresponds conceivably to the TM$_1^{(1, 0)}$ mode, which is blue-shifted by the sample tilt. 

For the D2-NAROLED device [Fig.~\ref{fig:SI-0deg-cuts}(b)], the low signal-to-noise ratio makes mode identification from the experimental data difficult. For DQ-NAROLED, the strongest peak appears near the TE$_0$ resonance, similar to the SL-NAROLED device with array period 370~nm.  

\end{document}